\documentclass[onecolumn,secnumarabic,amssymb, nobibnotes,nofootinbib,aps, prd,showpacs,showkeys,unsortedaddress,10pt]{revtex4-1}
\pdfoutput=1

\def\bmi {\begin{minipage}}
\def\emi {\end{minipage}}
\usepackage{appendix}
\usepackage{epsfig}
\usepackage{graphicx}
\usepackage{color}
\usepackage{soul}
\usepackage{ulem}
\usepackage{amssymb}
\usepackage{hyperref}
\usepackage{multirow}
\usepackage{float}

\def\ee{\end{equation}}
\def\be{\begin{equation}}
\def\eea{\end{eqnarray}}
\def\bea{\begin{eqnarray}}
\def\bal{\begin{align}}
\def\eal{\end{align}}

\def\be{\begin{equation}}
\def\ee{\end{equation}}
\def\bea{\begin{eqnarray}}
\def\eea{\end{eqnarray}}

\def\bef{\begin{figure}}

\def\eef{\end{figure}}
\def\bib{B\kern-.05em{I}\kern-.025em{B}\kern-.08em}
\def\btex{B\kern-.05em{I}\kern-.025em{B}\kern-.08em\TeX}
\def\RW{Rolf Wider\o e}
\def\BT{Bruno Touschek}
\def\Archiv{\it Archiv f\"ur Elektrotechnik}

\begin{document}
\title{Bruno Touschek, from Betatrons to Electron-positron Colliders}
\author{Carlo  Bernardini}
\email{carlo.bernardini@roma1.infn.it}
\affiliation{Physics  Department, University of Rome Sapienza\\
Rome, 00185, Italy}
\author{Giulia Pancheri}
\email{pancheri@lnf.infn.it}
\affiliation{INFN Frascati National Laboratory, Via E. Fermi 40\\
Frascati, I00044, Italy}
\author{ Claudio Pellegrini}
\email{pellegrini@physics.ucla.edu}
\affiliation{Department of Physics and Astronomy, University of California at Los Angeles, Los Angeles, California 90095\\SLAC National Accelerator Laboratory, Menlo Park, California, 94025}


\begin{center}
\vskip 2 cm
{\it Preprint of an invited article submitted for publication in \\
Reviews of Accelerator Science and Technology, Vol. 8.
http://www.worldscientific.com/worldscinet/rast \\
 \textcopyright  2015 by World Scientific Publishing Company }
 \end{center}
 
\begin{abstract}
Bruno Touschek's life as a physicist spanned the period from World War II to the 1970s. He was a key figure in the
developments of electron-positron colliders, storage rings, and gave important contributions to theoretical  high energy physics.
Storage rings, initially developed for high energy physics,
are being widely used in many countries as synchrotron radiation sources and are a tool for research in physics, chemistry, biology
environmental sciences and cultural heritage studies. We describe Touschek's life in Austria, where he was born, Germany, where he participated to the construction of a betatron during WWII, and Italy, where he proposed and led to completion  the first electron-positron storage ring in 1960, in Frascati.  
We highlight how his central European culture influenced his life style and work, and his main contributions to physics, such as the discovery of the {\it Touschek effect} and  beam instabilities in the larger storage ring ADONE. 
 
\end{abstract}


\maketitle
\section{Introduction}
This article  recalls the events and personalities leading to the birth of electron-positron colliders through  the life and works of  the Austrian physicist Bruno Touschek, who suggested, designed and brought to completion the first electron-positron storage ring accelerator, AdA, in Frascati, in 1960.  He was instrumental in bringing AdA to Orsay, in 1962, and, by discovering and explaining  what is now known as the {\it Touschek effect} \cite{Bernardini:1997sc}, opened the way to the extraordinary development of storage rings during  the second half of last century.  These novel particle accelerators gave many important contributions to high-energy physics and the development of the standard model of elementary particles. In addition, with the development of storage ring based synchrotron radiation sources generating very high brightness X ray beams, they have given us many new scientific contributions in biology, chemistry and physics, exploring matter at the atomic and molecular level.

 Touschek's major accomplishment lies with AdA, the electron positron storage ring, operational in Frascati in 1961. In a brief outline of the history of AdA, presented at the Accademia dei Lincei in Rome, on April 24th, 1974, Bruno Touschek wrote: {\it  The challenge of course consists in having the first machine in which particles which do not naturally live in the world which surrounds us can be kept and conserved.} The feasibility of electron positron collisions was fully demonstrated after AdA was transported from Frascati to Orsay at the Laboratoire de l'Acc\'el\'erateur Lin\'eaire (LAL). The experiments carried out at LAL, the discovery of the Touschek effect and the demostration, for the first time in the world, that electron positron collisions had taken place and had been observed,  opened the way to all future electron-positron colliders.\footnote{Comment by J. Ha\"issinski in the docu-film {\it Touschek with AdA in Orsay} by  E. Agapito, L. Bonolis and G.Pancheri, \textcopyright INFN2013} Touschek's stature in physics is not limited to AdA. His life long interest in electromagnetic radiation effects, mostly concerning the emission of multiple soft quanta in the interaction of charged particles, is an ulterior proof of his focus on how to extract physics from the accelerators he worked on. Touschek's interest, as he became a mature physicist, was always to develop accelerators as an exceptional tool for the exploration of matter.
 
To write this paper, we have used many sources, some of them primary and unpublished, made available to some of the authors by Touschek's family. We have also made wide use of secondary sources, available in institutional archives and in the current literature. Unless otherwise noted, dates are from Edoardo Amaldi's  biography, published as a Yellow Book by CERN \cite{Amaldi:1981be}.
Information about Wider\o e are from his biography, edited by Pedro Waloschek \cite{Waloschek:1994qp}.
 Other details of the war period come from Touschek's letters to his father, some of them quoted in a recent article by Bonolis and Pancheri  \cite{Bonolis:2011wa}, others still unpublished. These letters, preserved by Touschek's wife, Mrs. Elspeth Yonge, were kindly made available  to Luisa Bonolis and to one of the authors, G.P., and  cover the period 1934-52. They are addressed to his father and stepmother, as {\it Lieber Vater, liebe Mutsch} or, alternatively as {\it Liebe Eltern}. In what follows we shall generically indicate these letters as {\it letters to his family}.

 Other material about the war period and the work on the betatron can be found in a work   about betatron projects in Germany during the war by Pedro Waloshek \cite{waloschek2012deathrays}, and in  an article  about Wider\o e entitled {\it  Why is the Originator of the Science of Particle Accelerators so Neglected, Particularly in his Home Country?}, written by a Norwegian expert in radio-oncology \cite{Brustad:1998aa}. In addition to the above, the essential literature about Touschek and the beginning of storage rings includes refs. \cite{Isidori:1999bk,Bernardini:2004rp,Bonolis:2005as,Greco:2004np}.

 Finally, since all three authors of this article have worked with Bruno Touschek, we have added some personal anecdotes and memories.

\section{Early years in Vienna: 1921-1942}
Bruno Touschek  was born in Vienna on February 3rd, 1921. His father was an officer in the Austrian Army. The mother, Camilla Weltmann, belonged to a Jewish family, well known in the artistic circles of the Viennese Secession. His maternal uncle,  Oskar Weltmann, was a doctor and  painter, one of his maternal aunts would  marry Emanuel Josef  Margold, an architect member of the Vienna Werkst\"atte, assistant to Josef Hoffmann. 
 \begin{figure}[h]
 \centering
\resizebox{0.4 \textwidth}{!}{
 \includegraphics{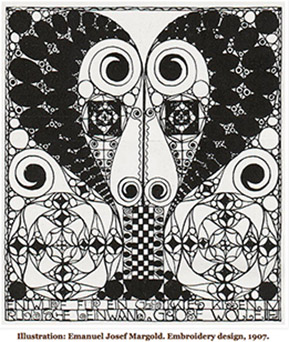}}
 \hspace{1.cm}
  \resizebox{0.41 \textwidth}{!}{
 \frame{ \includegraphics{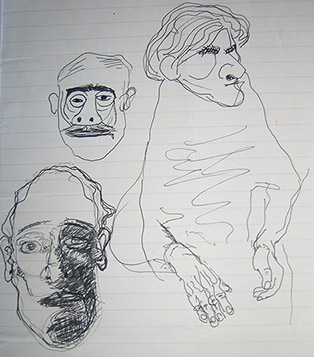}}}
 \caption{At left, an embroidery design  by Emanuel J.  Margold dated  1907 and,
 {at right,} a drawing by Bruno Touschek, probably 
 {dated} 1968.}
\label{fig:margold}
 \end{figure}
Another maternal aunt, Adele, nicknamed Ada,\footnote{This nickname  is confirmed by a letter written by Bruno to his family,  from Rome, and dated March 11, 1939, and it is of interest in connection to the  acronym given to the storage ring AdA, as seen later. } married an Italian businessman and lived in Rome.

Bruno lost his mother when he was a boy, in 1931, and remained close to his mother's family all his life. In particular, he often visited Rome and his aunt Adele during summer or school vacations. 
\begin{figure}[h]
\centering
\resizebox{0.3\textwidth}{!}{
\includegraphics{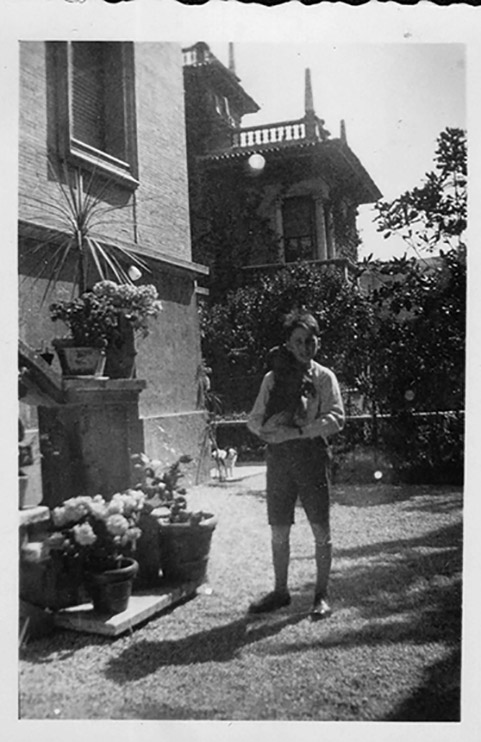}}
\hspace{1cm}
 \resizebox{0.6 \textwidth}{!}{
  \includegraphics{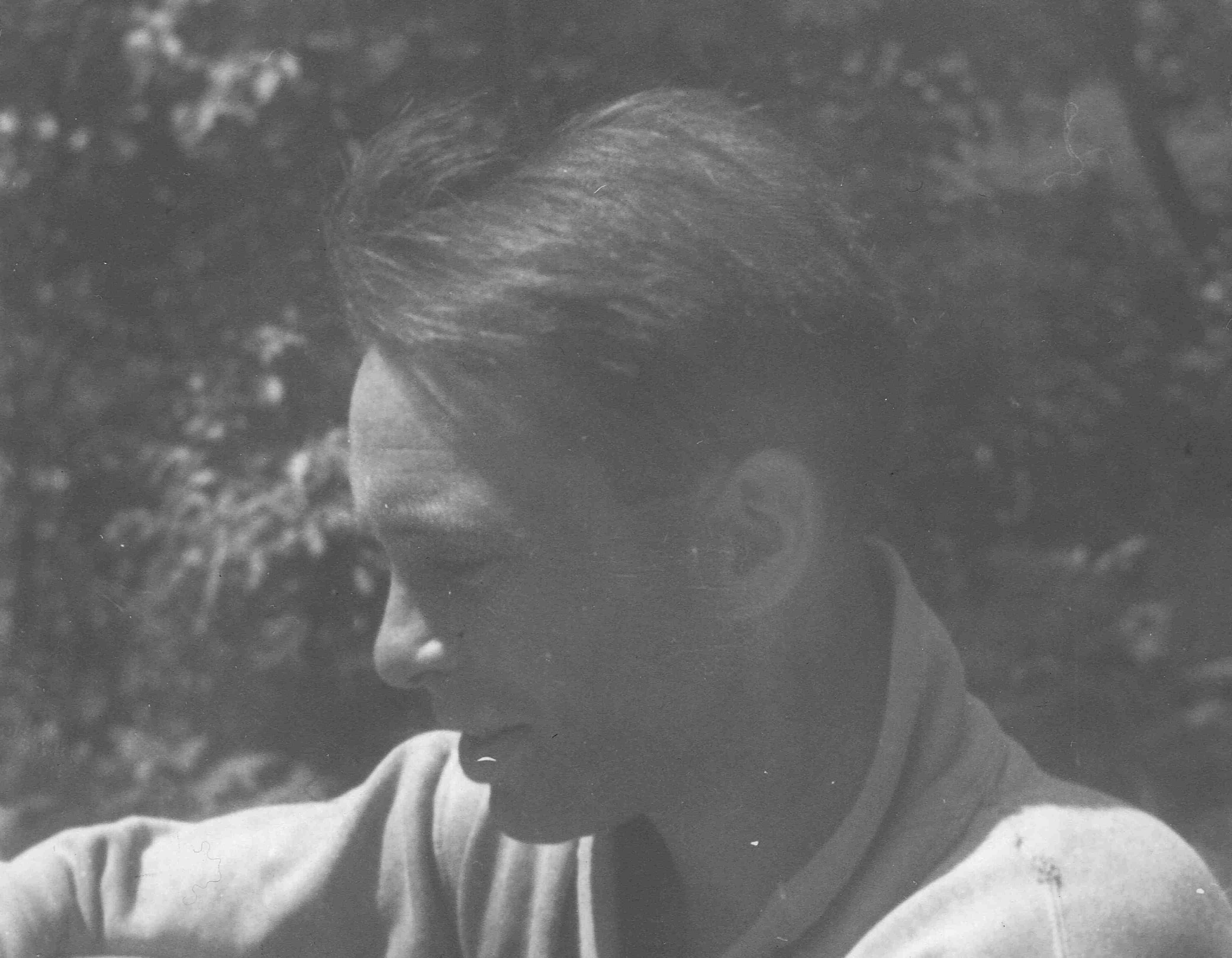}}
\caption{At left, Bruno Touschek  in 
{aunt Ada's}
 apartment, in  Rome. This photo was taken during one of his trips to Rome, probably  in 1936. 
 {At right, in later years, Touschek at his aunt's place, near Lake Albano.}}
\label{fig:BTrome}
\end{figure}
His last visit before the war took place, it appears, in Spring 1939, when he went to Italy  after passing his high school exam, {\it matura}, as a private student in March. The annexation of Austria to Germany, which had taken place one year earlier, had made difficult the life of Viennese Jews and he had to leave  the public gymnasium,  because of the restrictions imposed by the new authorities \cite{Amaldi:1981be}.\footnote{School records examined by Prof.  Mitter from Graz (private communication to G.P.) do not confirm this point, which is however present in Touschek's  recollections,  which were prepared  by Amaldi and checked with Bruno,  before his death.}
While in Rome, in Spring 1939, Bruno attended some classes at the University, in the engineering department, and started planning to study  chemistry at the University of Manchester, in England. From letters of this period to his father and stepmother,\footnote{In these letters, as well as in the whole correspondence of the period 1934-52 in possession of the Touschek family, Bruno often addresses his letters to his parents, as {\it Liebe Eltern} [Dear Parents], for instance as   early (after his mother's death) as in a letter dated 23rd July 1934. In other letters, for instance in  the  letter from Rome, dated March 11, 1939, he writes {\it Lieber Vati, liebe Mutsch!} } it appears that he was very hopeful and expecting the visa around the end of March. Nothing seems to have come out of this.

 In September 1939, the war broke out.  Bruno was back  in Vienna, and enrolled in physics at the University. This opportunity however did not last long: in June he was told he could not 
 register any more as a regular student.
 He  continued his education in physics, studying   on books borrowed from the University  library by a young assistant professor, Paul Urban. One of these books was the fundamental treatise  by Arnold Sommerfeld, {\it Atombau und Spektrallienen}, the bible of atomic theory.  Thus, through his very careful reading of this book, Touschek learnt quantum theory and entered in contact with Sommerfeld.  Thanks to the help and mediation of Paul Urban, a series of letters and at least one visit were exchanged between the young student Bruno and the great theoretical physicist in the fall of 1941.\footnote{The letters can be consulted at the Deutsches Museum in Munich, Germany. Communication courtesy of L. Bonolis.} Sommerfeld was clearly impressed by the youth and gave his blessings to the plan proposed by Urban and Touschek: to leave Vienna and go to continue studies in Germany, at the Universities  of Hamburg or Berlin, where Sommerfeld could recommend Bruno to colleagues he trusted,  as Paul Harteck \footnote{P. Harteck, together with M. von Laue, W. Heisenberg, C. F. von Weizs\"acker, were some  of the scientists held prisoner by the British government in Farm Hall, after the war.}, Von Laue, Jensen. In the early months of 1942 the plan came into being and on February 26th   1942, Bruno wrote to his parents from Munich, where he had stopped to meet Sommerfeld, who gave him   a letter for  Paul Harteck, then teaching at the University of Hamburg.   
\section{Germany 1942-1947}
\begin{figure}[h]
\centering
\resizebox{0.4\textwidth}{!}{
\frame{\includegraphics{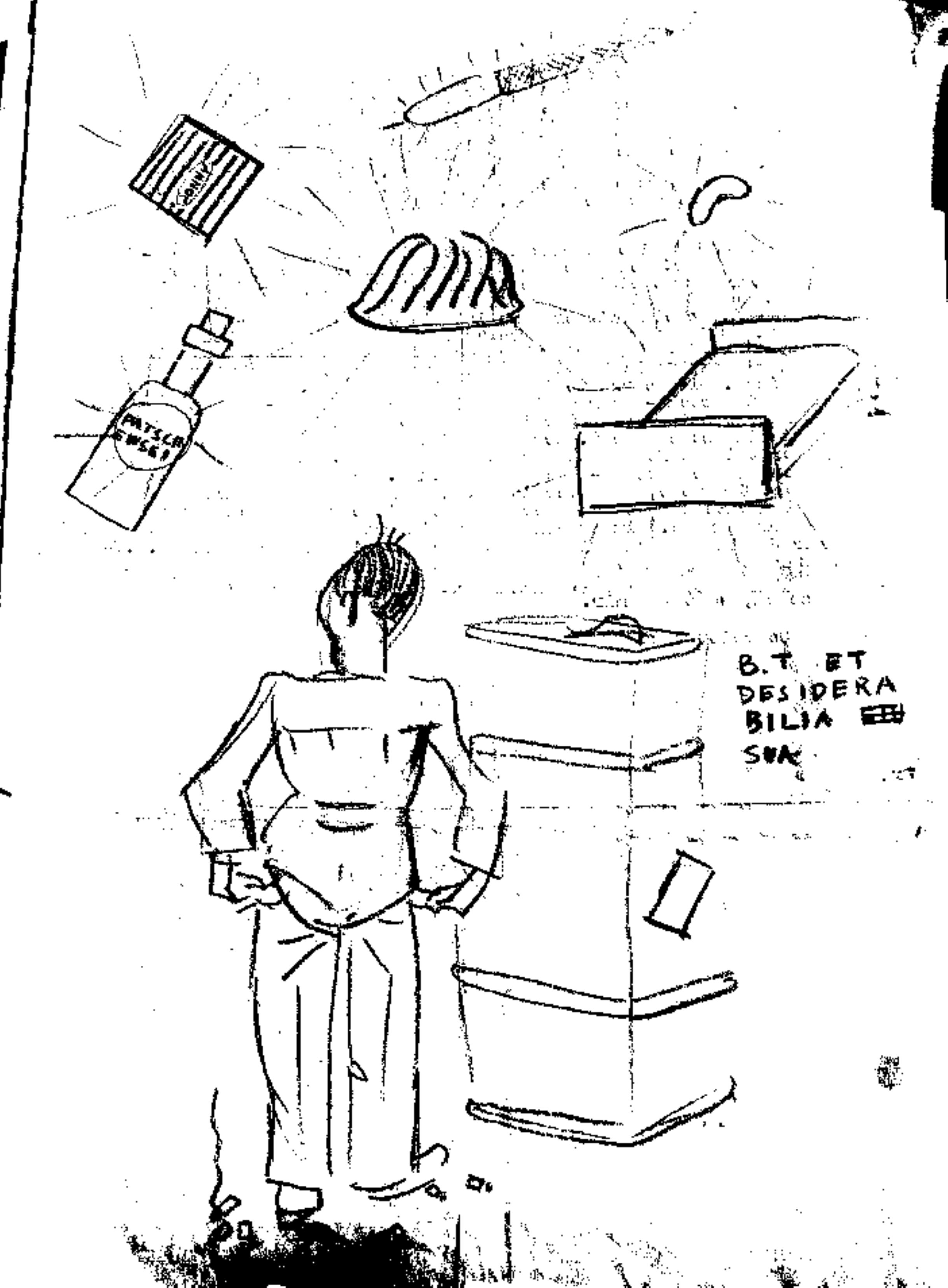}}}
\caption{A graphical list of Bruno's desires, 
{extracted from } a letter   to his parents, 
{ Hamburg, 1942.} }
\label{fig:1943desiderabiliasua}
\end{figure}

Touschek arrived in Hamburg in March 1942, and spent  all the war years between this city and Berlin. He attended classes in Hamburg, but often travelled to Berlin, to follow other courses. He  was always looking for some way to earn money for his  living. His letters home in this period reflect his loneliness and difficulties with   money, suffering from  cold, lack of food and  no comfortable room, as graphically shown in a letter to the family,  in Fig. ~\ref{fig:1943desiderabiliasua}.

In Hamburg he attended, among others,  the classes of Paul Harteck, in Berlin those of Max von Laue. 
 Sometime in 1942, on a train to Berlin, he met a young woman, herself half Jewish like Bruno, and through her he was able to secure a job in Berlin. This job  would bring him in contact   with Rolf Wider\o e and set the road leading to    AdA,  many years later. He worked  in a small factory, Opta-Radio, whose name had been changed for political  reasons from the former  (too much Jewish sounding) L\"owe-radio.  In Germany, at the time,  there were many half Jewish scientists and engineers,  tolerated by the
 {Nazi} regime when involved in useful technical work. They  supported and helped each other to  secure some useful work in technical or scientific concerns belonging to them or their friends \cite{waloschek2012deathrays}.
 At L\"owe-Opta, Bruno  spoke  with Dr. Egerer, who was chief Editor of the scientific journal {\Archiv}.  At the time,  he was quite young, but already had very good credentials,  as he had read  and studied Sommerfeld's book, among others, and was following advanced courses at the University.  At the end of 1942, in November,  Egerer offered him a position and  Touschek accepted it.  One of his duties was to be an assistant to the editorial work for   {\Archiv}.  It thus happened that, in early 1943, Touschek came across an article submitted to  {\it Archiv } by a Norwegian engineer.  

In the next subsection, we shall describe the events that  in 1943 brought together the Austrian Bruno Touschek and the Norwegian engineer Rolf Wider\o e, who in 1928 had  first proposed   the betatron to accelerate electrons to high energies. But to fully understand this, we must first  step back to 1940.

\subsection{Touschek's  war years and  Wider\o e's betatron}
Rolf Wider\o e had developed the betatron concept and analyzed it mathematically in 1928, in an article summarizing his PhD thesis at the University of Karlsruhe \cite{Wideroe:1928}. His  attempts to make a working electron accelerator were not successful and, after some time, he abandoned this research and went to work for the Norwegian subsidiary   of Brown Bovery, in Oslo. However, his article was read by Lawrence, then  building his  first  cyclotron and subsequently read and cited by Donald Kerst \cite{Kerst:1940zz,Kerst:1941zz}, who built the first betatron, which he called the induction accelerator. Kerst's work and the following article jointly with Serber \cite{Kerst:1941serber} were published in  {\it The Physical Review} in 1940 \footnote{In Kerst's article,  Wider\o e's paper is wrongly dated as been published in 1938. In fact, it was published in 1928.} and 1941 respectively. The 1940 issue was the last to reach Norway, which had been occupied by Germany since April 1940. Sometimes in 1941, this issue  reached Trondheim, where the physicist Roald Tangen  read it with great interest and decided to present the latest results on accelerator developments,  including Kerst's work,   in  a talk    at   the Physics Association, in Oslo, in December. At this seminar Roald Tangen illustrated  Kerst's results, and commented on the first reference in the Physical Review paper, to  someone with  a Norwegian name, some Wider\o e. \RW , always interested in accelerator developments, was attending the Conference.   Clearly his interest was renewed, and  he saw that the betatron principle worked and others were already developing  it. Thus,  in the months to follow, he prepared an article about the {\it Die Strahlentransformator}, which he submitted to  {\it Archiv}  on September 15, 1942.

The article proposed to build a 15 Mev betatron, the first of such energy to be built in Europe. Because, as we shall discuss below, the proposal contained in  this article was actually presented and then  accepted by the German Ministry of Aviation, all the information about this article became later classified, and the reconstruction of the events, proposed here, is based on information contained in one of Touschek's letters to his family during the year 1943, and from what Wider\o e \ himself writes in his autobiography. 

 In a letter to his parents, dated February 15, 1943, Bruno mentions  having  read a {\it crazy } article. 
 A translation of this letter is given in \cite{Bonolis:2011wa}. 
In this letter, in addition to comments about some calculation of electron orbits missing proper relativistic corrections,  Egerer is described as being excited about possible plans for building something of interest to the War Ministry and to Heisenberg.  Egerer, who had connections with the  Ministry of Aviation of the Reich,  Reichsluftfahrtministerium (RLM), must have presented the idea to the military and made the officers of the RLM interested in \RW. In fact, sometime in the Spring 1943, \RW , in Oslo, was approached  by German officers and asked if he would be interested in building a betatron and, upon his affirmative answer, was flown to Berlin the following week, presumably to present his project.\footnote{Why \RW\ accepted and how this was discussed after the war by an investigation Committee in Norway, is discussed both in Wider\o e's autobiography and in \cite{Brustad:1998aa}.} Other betatron projects existed in Gemany at that time, and were discussed at the RLM level \cite{waloschek2012deathrays}, and it appears that Wider\o e's was considered the most reliable (after all it had been \RW \ who had first proposed the theory in 1928) and, from another of \BT's letters, this time dated June 1943, we learn that  the project started and that \BT \ was part of the team. Because of the secrecy surrounding the project, Touschek's letters are not very explicit   concerning his relationship with \RW, Touschek referring  to him as W. or {\it Mein  Norweger}. From what \RW\  himself wrote to Amaldi after Touschek's death, Touschek and Wider\o e  appear to have exchanged a number of letters  to correct some theoretical calculations, mostly about the relativistic treatment of the electron's orbit. From Touschek's letters to his family, it seems  that he encountered some resistance on Wider\o e's part. Finally in June 1943, Touschek  writes to his parents: ``I have beaten my Norwegian on all fronts." 

According to the already cited Wider\o e's autobiography \cite{Waloschek:1994qp}, the idea of colliding oppositely charged particles  was conceived by \RW\ in the summer months  following  the approval  of the project. Bruno Touschek recalls  conversations with \RW\ in the notes he wrote  to  prepare  the proposal of AdA presented  to the Frascati Laboratory on March 7th, 1960.    He writes:     {\it The following is a very sketchy proposal \dots  Let me first explain why a storage ring is a very important instrument, particularly when fed with electrons and positrons.\dots The first suggestion to use crossed beams I have heard during the war from Wider\o e \dots }. We will see in subsection~\ref{ssec:AdAprop} how  AdA's  proposal came about and was received. Here we will  continue with Touschek's war years, and the period before coming to Italy.  

\begin{figure}
\centering
\resizebox{0.8\textwidth}{!}{
\frame{\includegraphics{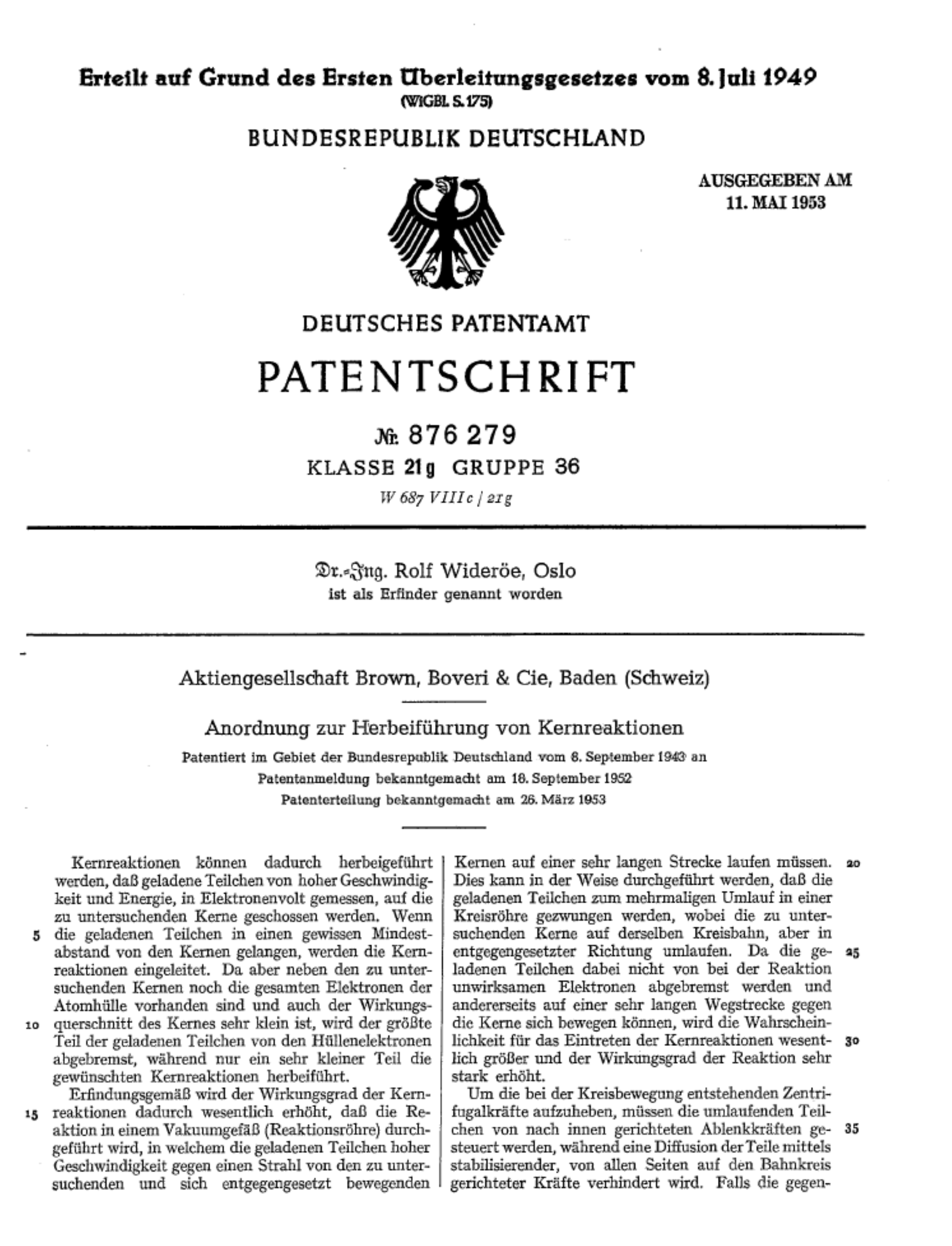}}
\hspace{+5cm}\frame{ \includegraphics{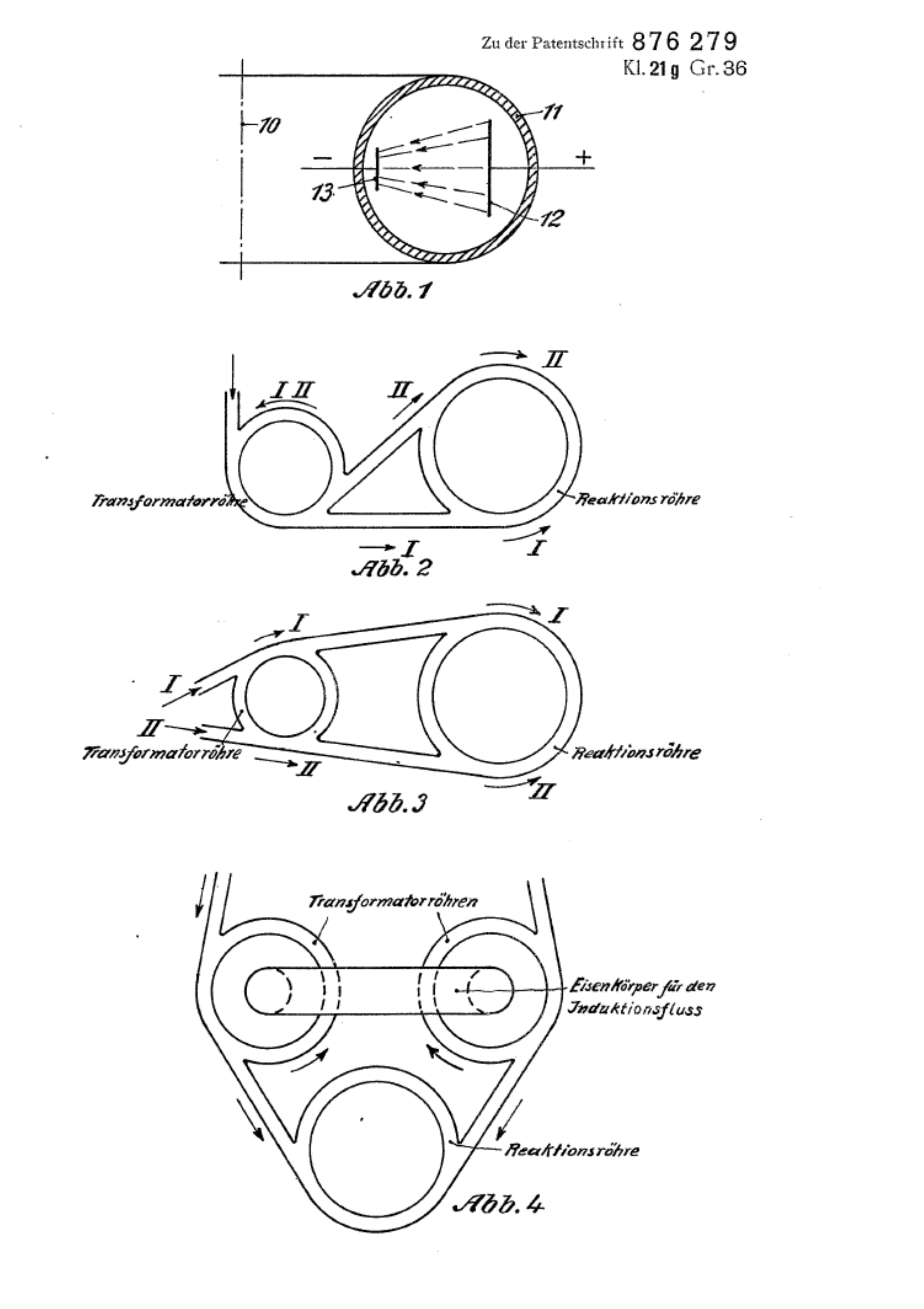}}}
\caption{At left we show the first page of the patent submitted by Wider\o e on September 8, 1943 for collisions between oppositely charged particles, at right a detail of the proposed set-up, from \cite{Waloschek:1994qp}.}
\label{fig:widpatent}
\end{figure}

On September 8th, 1943, a patent  was submitted by  \RW \  for an accelerator with oppositely charged particles colliding in a ring.\footnote{Patent n.  876279, courtesy of Luisa Bonolis.}  We show in Fig.~\ref{fig:widpatent} some details of this patent, including a possible set-up. 
\BT\ did not, in 1943, think much of Wider\o e's idea, which he dismissed as trivial, but years later, when   accelerators became the central tool for particle physics research and new particles were discovered,  the idea came back to him, and became the idea of the century. At the end of this year, the betatron project became fully classified, with the consequence that  Wider\o e'article for a 15 MeV betatron, the one which initiated the project, was never actually published.\footnote{In \RW's autobiography, the article is declared as published, and a bibliographic reference is given. However no such article can be found in the collection of {\Archiv} \ for the years 1942 to 1945. 
 Copy of \RW's article exists in  proofs form and has been obtained by Bonolis and Pancheri, courtesy of Pedro Waloschek.}

\begin{figure}[h]{!}
 \centering
 \hspace{-1cm}
 \resizebox{0.36\textwidth}{!}{ 
\frame{\includegraphics{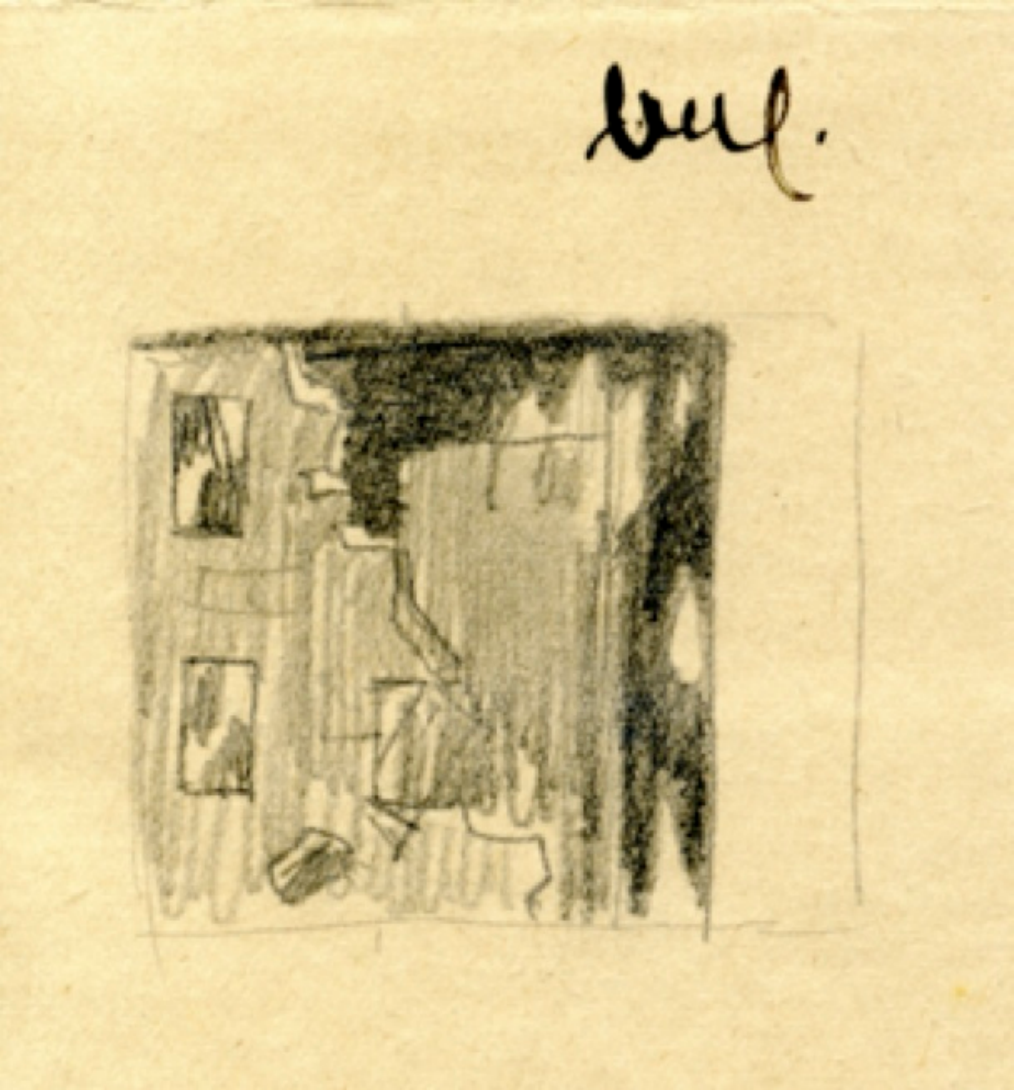}}}
\hspace{1cm} 
\resizebox{0.52\textwidth}{!}{ 
\frame{\includegraphics{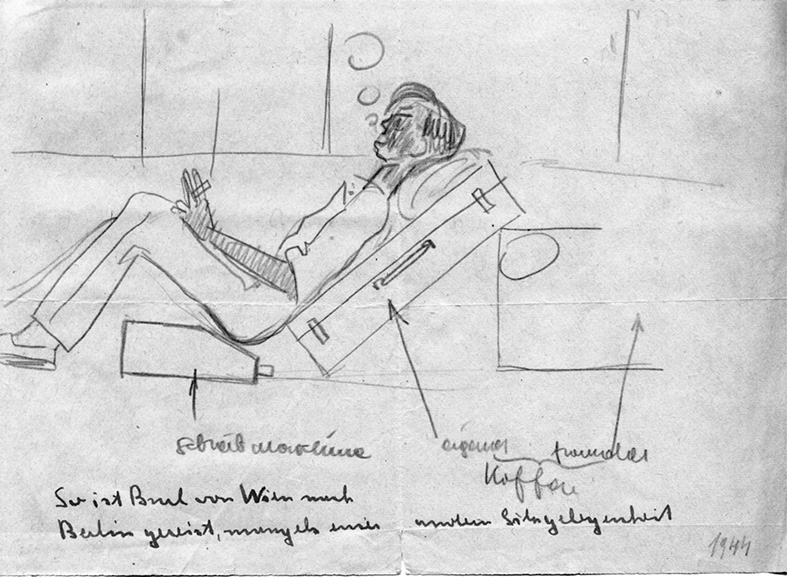}}}
 \caption{At left, we reproduce  a drawing of destroyed buildings in Berlin from a letter dated November 23rd, 1943. At right, a drawing showing Bruno during a train travel from Vienna to Berlin, in 1944, from \cite{Bonolis:2011wa}. Courtesy of Touschek's family.}
 \label{fig:1943-44}
 \end{figure}

At the end of 1943, we therefore find \BT \ fully employed in the secret betatron project. He knows the danger of being involved in this, and writes to his parents: ``On this date, I have signed my death warrant\dots". He had clearly understood that by signing a contract to work for RLM, his life would become known to the authorities and his Jewish origin could not escape the Gestapo's attention. However, there is no mention of this later aspect, except that he often refers to the Labor Office, OT, which controlled the work of foreigners, or non-Aryan personnel. In these last months of 1943, in the meanwhile, the situation in Germany worsened. Berlin became the target of extremely heavy air attacks by the allied fire-bombings. These raids are recalled by \BT\  in two letters to his parents, one on  November 23rd, one on December 16th, 1943. As was often the case, his letters to the family included little drawings,  comments to his tales of hard life. In  Fig.~ \ref{fig:1943-44} we show, on the left,  one such drawing, illustrating the destruction brought in by air raids. 

The year 1944 sees Bruno  working on the betatron, providing theoretical calculation, on relativistic aspects and on radiation damping, a subject which he was concerned about all along his scientific life, as detailed later. He occasionally visited his family in Vienna, and in the right panel of Fig. ~\ref{fig:1943-44} we show a drawing illustrating  his travels from Vienna to Berlin. 
He was also following classes in Berlin and Hamburg, where he finally moved in 1944, upon  Wider\o e's request. The betatron was being built near Hamburg, at the C.H.F. M\"uller factory,
the R\"ontgenr\"openwork (X-ray tube workshop),   and it was more practical to be there, rather than continuing to move back and forth.
\ 

At the end of 1944, the betatron was operational. Touschek went to visit his parents in Vienna in the fall 1944, sending them a letter with the drawing  shown in Fig.~\ref{fig:1943-44}. As 1945 started, the war was near  its end. In March, the German authorities decided that the betatron needed to be put in safety outside Hamburg, which was obviously a target for the Allied forces marching towards Hamburg and Berlin. An abandoned factory north of Hamburg, belonging to one of the members of the betatron group was found, and Touschek with Wider\o e moved the betatron there. On Wednesday, March 16th , 1945, Touschek writes to his parents from Kellinghusen, sitting on the veranda of the house with Wider\o e: ``It is Wednesday, and I am exhausted".

As soon as the betatron was stored in temporary safety, Touschek was arrested by the Gestapo. The coincidence of the events, whose precise chronology can be found in \cite{Bonolis:2011wa}, indicates that the Gestapo must have been  aware all along of Bruno's  Jewish origin, and that once the task was accomplished - the construction of the betatron - he could well follow the fate of all other Jews in Germany. He was put in jail,\footnote{This was the Fuhlsb\"uttel prison, known for suffering and death,  a temporary jail for prisoners awaiting to be sent to concentration camp or other jails.  }  without knowing why, being told that he was accused of espionage, with other sick prisoners, mostly Jews, he notes, no cigarettes of course, well aware that he risked death, just when he had expected to be able to go back to Vienna. But he had been working for the RLM, and after one week in jail, during which he meditated suicide, his friends, certainly Wider\o e, did come to bring some relief, although they could not get him free. 

The events of the last months of the war and how Bruno escaped from death in the Kiel concentration camp, are described in two letters sent by Touschek to his parents, after the war was over, the first one dated 
{J}une 1945 arriving on October 22nd (as carefully noted by Touschek's father on the frontispiece of the letter) and the second arriving in November. Both letters are published, in English translation in \cite{Bonolis:2011wa}.  In these, he describes the weeks in jail and recounts the dramatic days in April 1945. Around April 11 or 12, as the Allied forces were approaching Hamburg, the prisoners of the Fuhlsb\"uttel jail, all 200 of them in a long line with SS guards at rear  and the front, as  Touschek writes,   were  marched  to the Kiel concentration camp, 40 kilometers away to the North. On the way, Touschek, sick, apparently with a load of books, or other belongings, fell to the side of the road, and a guard shot at him. Whether by purpose or by accident, Touschek was not seriously hurt: he lost consciousness, after one of the bullets pierced his ear, and  was left  behind,  apparently to die. The accident saved his life. He then spent time in some hospital, and  again imprisoned.  But the war was ending, and soon he was freed and joined the English forces, as an interpreter, and even resumed his contact with the betatron group, except for Wider\o e, who  returned to Norway in April. In the letters, already mentioned, to his parents, dated October and November 1945, Touschek notes that nobody seemed to know where Wider\o e \  was:  indeed, upon his return to Norway, he had been  accused of collaborating with the Nazis and was put in prison for 47 days. He went through an  investigation, at the end of which he had  to pay a large sum of money, as reparation. The  details are described in \cite{Brustad:1998aa}.   
\section{Formation as a theoretical physicist: from  Glasgow to Italy }
After the war, Touschek  was finally able to complete his study and obtain the formal degrees.
He went to G\"ottingen, where he obtained his diploma with a thesis on the theory of the betatron,  and then to Glasgow, where he was awarded his doctorate in physics. Later in his life, he would wonder why he had  not been able to remain in Germany, with Heisenberg's group. He was not happy in Glasgow. Mostly he resented the weather, and, perhaps the rather different ways of the region. An incident, which highlights the young 
{man's} high-strung disposition is told by  P.I. Dee to Amaldi: during one summer, probably 1949, Touschek had left Glasgow to help the harvest in the North. When he returned he discovered to his utter dismay that the landlady had changed the curtains in his room, without asking for his permission. 	The feeling of injustice and prevarication, even on such irrelevant issue, were too strong for the young man and he could not accept the change, and soon moved out.

In Glasgow Touschek worked on a synchrotron project. Remaining in Glasgow  after his doctorate, he  established a friendship with Walter Thirring, also a young post-doc there. Walter   was the son of Hans Thirring, whose lectures at University of Vienna Touschek had sometimes followed, when traveling {\it incognito} from Germany back to Austria, as we have seen in the drawing of Fig. ~\ref{fig:1943-44}.
With Thirring, Touschek wrote an interesting paper on a topic of great theoretical interest, albeit not so in practical terms at the time, the so called "infrared catastrophe", discussed in a pre-war paper by F. Bloch and A. Nordsieck \cite{Bloch:1937pw}.  Bloch and Nordsieck  had discussed  radiative emission in the scattering of charged particles and  the well known problem of the divergence accompanying single light quanta when their frequency went to zero. They had shown  that     the probability of emission of an  infinite number of quanta was  always to be finite. Considering this to be one problem admitting of a closed form solution in field theory, and thus of great potential interest,   together, Touschek and Thirring, reformulated it  in covariant form \cite{Thirring:1951cz}, and with this  Touschek laid the foundations for some of his future work for infrared radiative corrections to colliding beam experiments. 

In 1952, Edoardo Amaldi offered to  Bruno Touschek   a position with INFN, the Italian Institute for Nuclear Physics, newly established  to foster nuclear and particle physics research in Italy. The position  was at the University of Rome.  Touschek  accepted  it and moved to Rome. One of his maternal aunts,  aunt Ada, lived here and  he   had often spent his vacations in Rome,  before the war.   Fig. ~\ref{fig:BTrome} refers to one such vacation and   Bruno must have felt   this move to Italy as  a return  to the happy days before the Anschluss. He enjoyed the food and the climate, and found a very congenial scientific environment. Physics research was growing and expanding in Italy and Rome,  
following an old dream of Enrico Fermi, who had proposed to create a national laboratory for nuclear research and the construction of a particle accelerator since the 1930s, before emigrating to the United States.
Thus, it  is quite possible that Amaldi's interest in Touschek was arisen by Touschek having done his diploma thesis on the betatron and his PhD thesis on the synchrotron. 

 In the years between 1952 and 1959, Touschek was mostly engaged in theoretical particle physics work, collaborating with various colleagues, on problems related to symmetry properties and transformations, such as time inversion and continuous chiral symmetry, which he was the first to introduce in a 1957 paper on parity conservation and the mass of the neutrino \cite{Touschek:1957nn}. His  article with Marcello Cini on the relativistic limit of the theory of spin 1/2 particles\cite{Cini:1958aa}, written during this time,  is still studied today.

\begin{figure}
\centering
\resizebox{0.47\textwidth}{!}{
{%
\setlength{\fboxsep}{0pt}%
\setlength{\fboxrule}{1pt}%
\fbox{
\includegraphics{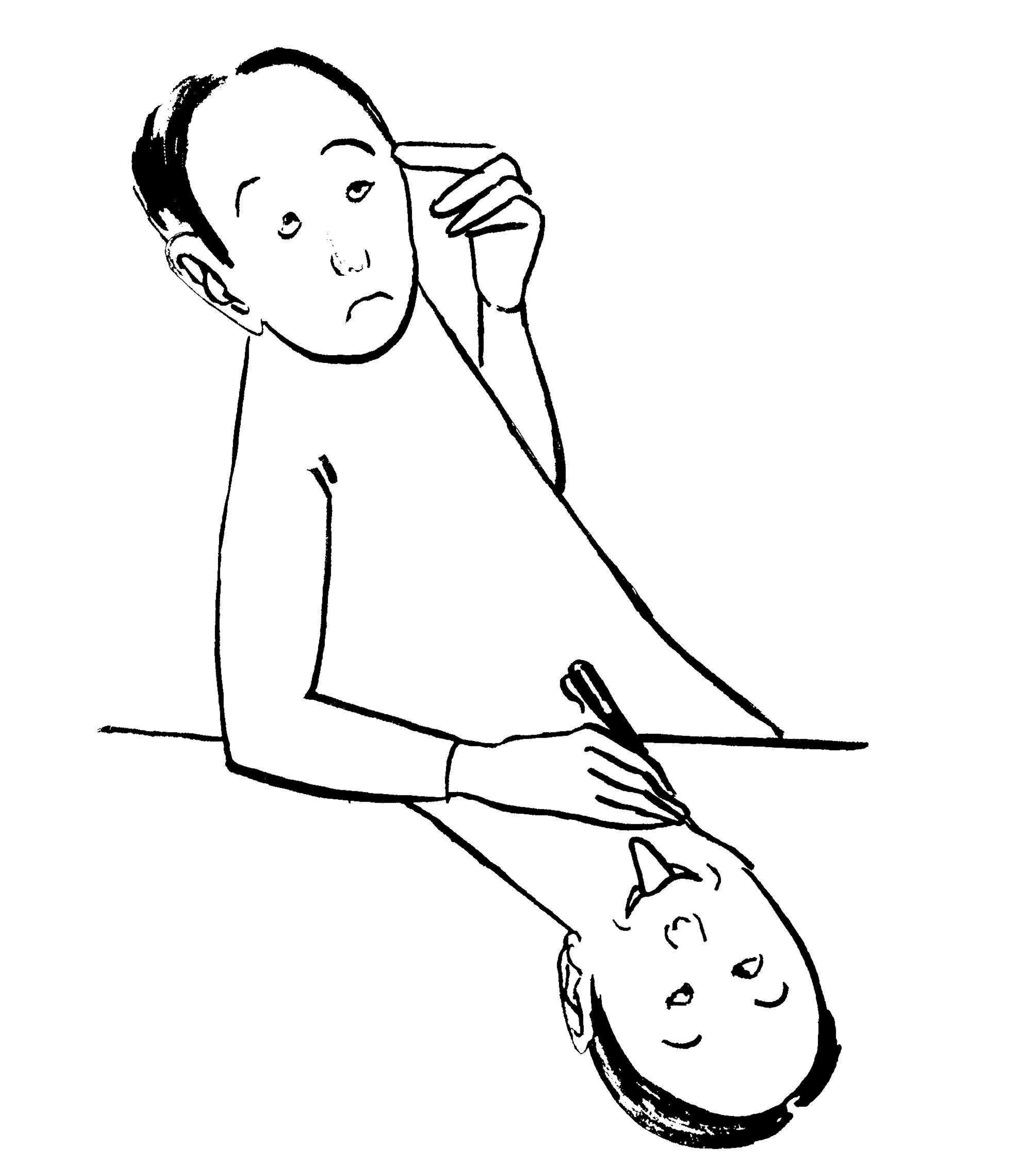}}}}
\hspace{1cm}
\resizebox{0.417\textwidth}{!}{
\includegraphics{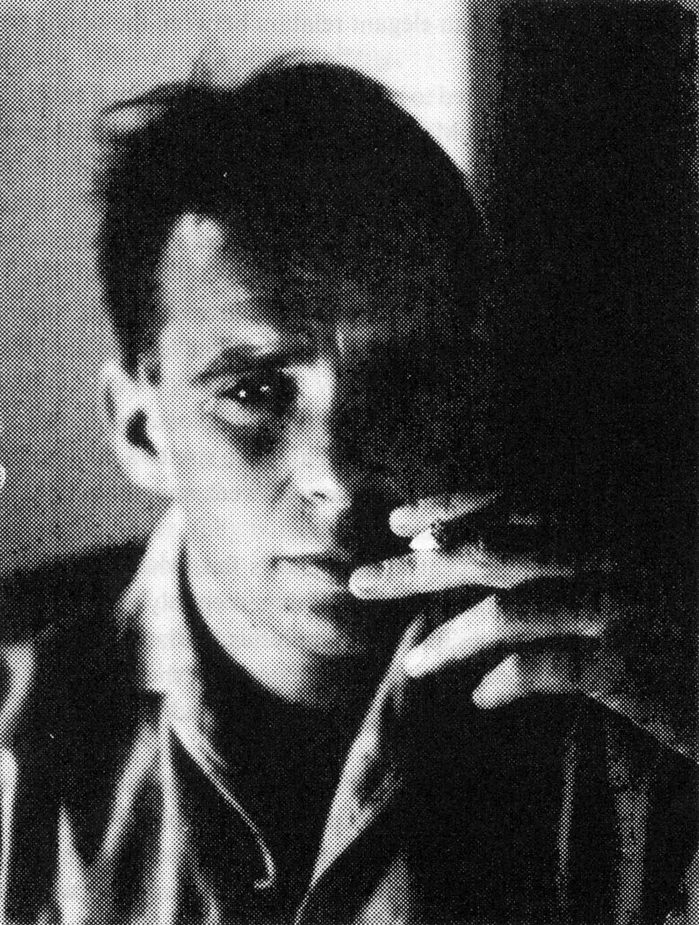}}
\caption{
{At left, a representation of parity violation through a portrait of Bruno's good friend, the 1957 Nobel prize winner  T. D. Lee. At right,   the well known  photograph of Bruno Touschek,  dated  1955 in \cite{Amaldi:1981be}.}}
\label{fig:sigaretta}
\end{figure}

A complete and extensive description of Touschek's theoretical work in this period  can be found in  \cite{Amaldi:1981be}. In this work, we also find  the well known photograph of Touschek shown  in Fig. ~\ref{fig:sigaretta},  as well as Touschek's  caricature of T. D. Lee, Nobel prize winner in 1957  for  the discovery of non conservation of parity, and  a life long  friend of Bruno Touschek. The discovery of parity violation in 
interactions observed in $\beta$-decay in 1956 and understood as such by T.D.Lee and C.N. Yang in 1956 \cite{TDlee:1956}  contributed to a shift in interest from pure QED studies to particle physics, where   the discovery of  
 new  particles, such as  kaons and vector mesons, were changing the scientific picture. A clear need to study the properties of these newly found particles emerged, and, with it, the need to   develop the new accelerators to do it.  

 \section{ The electro-syncrotron, and AdA}
 While Touschek was engaged in theoretical work, the Italian National Laboratory  was funded in Frascati,  and work to construct   an electron synchrotron started in Frascati in 1954, under the direction of Giorgio Salvini, who had done his thesis on the Kerst and Serber article, in 1941. At the end of 1959, the Frascati electron-synchrotron was  tested at full energy (1.1 GeV): everything worked as programmed. A general sense of success pervaded the staff, both operators and experimenters: injection was not a simple matter because of a last moment change; a van de Graaf
Accelerator was installed instead of  the original Cockcroft-Walton. The proton synchrotron of the Conseil Europ\'een pour la Recherche Nucl\'eaire (CERN) had also started operation and among the Italian physicists there was a dilemma: is it better
to work at Geneva or Frascati? The preference for Geneva was clearly
motivated: protons are much more generous than electrons in producing
interesting final states, leading to  faster production of new results with protons. Electrons gave pale and 
 disappointing beams, because the fine structure constant  dominates the cross sections. 
However, as Bruno Touschek had already noticed, proton reactions induced
confusion like hooligans in a market. Bruno was deeply involved with Quantum Electrodynamics (QED),
but not as most other  physicists working on it. The Feynman pictorial
representation of elementary processes gave a powerful language to
classify events, Bruno was very familiar with it and was attracted by such
problems as the infrared catastrophe, a serious correction to be introduced in
the analysis of experimental results. On the contrary he was not interested
in hunting  the breakdown of quantum electrodynamics, in which, after the introduction of the Ward
identities in 1950, most theoreticians were dreaming  to discover the future of
elementary particle physics. A positive friendship began  after a visit
of Robert Hofstadter, from Stanford, to Frascati, when the role of form-factors
became the center of speculations. 

Towards the late 1950's, Bruno was asking himself and the
colleagues (such as  Raul Gatto and Nicola Cabibbo): Òwhich are the most
important signals elucidating the elementary particle world?Ó. 
At that time his curiosity had started to  address   matter-antimatter annihilation, 
particularly $e^+ e^-$. Actually, in many and frequent discussions, Cabibbo and
Gatto were then pushed by Bruno to study the systematic classification of
annihilation processes having a broad variety of possible final states. This classification work led to a paper, published in {\it The Physical Review} \cite{Cabibbo:1961sz},  that was called  the ``Bible"  by the Frascati physicists to stress its basic importance.

In addition to a brilliant  group  of young physicists and engineers, it is important to note that the Laboratory could also count, at the time (end of 1959), upon  an exceptional team of technicians, with experience on extreme vacuum and radio frequency electronics, and who had learnt their art through  building  the synchrotron.

\subsection{Proposing electron-positron colliders}\label{ssec:AdAprop}
During the 1950's there were many discussions in Europe about the future roads to accelerator developments and the best way to access new energy frontiers for elementary particle physics. In a { 1956  meeting} at CERN, Giorgio Salvini asked  ``why not try electron-positron collisions?", but nobody knew  at the time  how to store positrons and make them collide with electrons. The concept of center-of-mass collisions was clear in the mid-fifties, also for the case of proton-proton collisions, but many concurring reasons made it difficult to extend it to matter-antimatter collisions. As Touschek said, a few years later, ``The challenge is to have enough anti-matter in the laboratory", in addition the physics to be explored with electron and positrons was  still limited to studies of Quantum Electrodynamics and the form factors. 

 In the second half of the 1950's American scientists
 began to   think about  tangent-rings  accelerators,  
and plans to construct an electron-electron collider started with a joint Princeton-Stanford project \cite{O'Neill:1958gk}. At the  time, but mostly unknown in the West because of Soviet secrecy surrounding all scientific activities, also  in the Soviet Union  scientists started considering   center of mass collisions.
 Activity for building an electron-electron storage ring  had   started at the end of 1958  under the direction of G.I. Budker, 
at the newly organized Institute for Nuclear Physics \cite{Baier:2006ye}.\footnote{This institute was  the Siberian Branch of the Academy of Science of the USSR. In 1957 the government had decided to create a large  scientific center in Novosibirsk and in 1958, the INP was formally created. Transfer of staff from Moscow to Novosibirsk took place in 1961 \cite{Baier:2006ye}.  }However, according to the theoretical physicist V.N. Baier \cite{Baier:2006ye},  $e^-e^-$ collisions were not considered interesting enough by the great theoretical physicists I. Ya. Pomeranchuk, who visited the new laboratory in October 1959.  Baier writes: {\it Pomeranchuk was not in raptures concerning the discussion and no support to the project was expressed.} Following  Pomeranchuk's visit,  the alternative idea of electrons versus  positrons  sprang up and, as Baier recalls,   on October 28, 1959, the  actual design of an electron-positron collider began.
At the beginning, Baier writes, there was not enough support from many famous members of  the Soviet Academy of Science   to move rapidly  and  effectively in that direction.
The project  did start in earnest in 1961 after   the complete  transfer of the   institute staff from Moscow to  Novosibirsk, in the {\it wild East}  (as Baier jokingly calls it),  and  after news of  AdA's results 
reached Novosibirsk.

The  year 1959 signalled a turning point  for particle physics in Europe: in  1959, the Frascati electron synchrotron started operation, followed shortly after by the Orsay Linear Accelerator, and, towards the end of the year, by the  CERN  proton synchrotron.  Within Europe, and in  Italy, there were frequent exchanges with American colleagues.  A visit and seminar by Wolfgang Panofsky in  Fall 1959  is said to have ignited the spark which led to AdA and the revolution in particle accelerators that  followed it. According to Nicola Cabibbo,\footnote {Comment  in  the docu-film {\it Bruno Touschek  and the Art of Physics} by E. Agapito and L. Bonolis, \textcopyright INFN 2003.} at the end of a seminar  by Panofsky, then 
Director of the High Energy Physics Laboratory of Stanford University,\footnote{ 
{L}ater known as the Hansen Experimental Physics 
{Laboratory.}}
  Touschek said: ``Why not try electrons against positrons?".\footnote{The exact date of the seminar is not given, but the list of seminars held at Frascati National Laboratories, includes a seminar by Panofsky on October 26, 1959, entitled ``On the two-miles Linear accelerator".}

It appears that things moved rapidly after this. Discussions at the University of  Rome and in Frascati became intense. Nicola Cabibbo, a young researcher at  Frascati in the years 1959-60,   and Raoul Gatto, at the University of Rome, started thinking of  what could be measured in electron-positron collisions and, on February 17th, 1960,  submitted a short paper to Physical Review Letters \cite{Cabibbo:1960zza}. 
 A few month earlier the idea of studying the pion form factor in the time-like region through the reaction $\pi^+ \pi^-\rightarrow N{\bar N}$ had been proposed by Frazer and Fulco \cite{Frazer:1959gy}
 and this inspired Gatto and Cabibbo to turn the reaction around and study  the pion form factor  with $e^+e^-$ rather than nucleon-antinucleon \footnote{In their paper Cabibbo and Gatto acknowledge their debt to Panofsky for ``a stimulating seminar on the possibilities of colliding beam collisions". }. From Touschek's notes dated February 18, 1960, we see that his attention goes beyond the studies of the pion form factor to  a list  of annihilation processess,  $e^+e^- \rightarrow \pi^+\pi^-, \ \mu^+\mu^-, \ \gamma \gamma$. Later, in his November 1960 proposal for building ADONE, he would add $n{\bar n}$.
  When a meeting was called in Frascati, on February 17th, 1960, to identify future programs for the Laboratories, Bruno Touschek brought up a proposal to store electrons and positrons and have them collide head-on. He proposed to use the synchrotron as a storage ring. He jokingly justified this proposal by writing:  ``Italy is a poor country and can hardly  afford two rings, thus let us use the synchrotron, which has already been built, as a single ring".

One of us, C.B., vividly remembers those days.  At the end of 1959, the Frascati 1.1 GeV electron synchrotron was ready to start experiments, even if the injection in the accelerator was giving some trouble because of low field irregularities disturbing injection. Many physicists used to 
 meet  in a little seminar room and discuss opportunities around a blackboard. Everybody was there, including Giorgio Salvini (then director of the Laboratory and alwys very active), Al Silvermann, theorists like Gatto, Cabibbo, Putzolu,  and many others. Bruno Touschek was among them, listening and commenting.  He was disappointed by the insistence on Òpion photoproductionÓ: undoubtedly new particles were most attractive, vector mesons in particular. Sakurai and vector dominance were not too far away; the $\eta$-particle was in the air.
In February 1960 Bruno declared that such routine work was not requiring particular skill because it consisted in using inefficient electrons, in place of much more efficient protons, to repeat pion physics (he started calling protons hooligans as compared to electron girls). Salvini, Querzoli, Silvermann and one of the authors of this paper (C. B.) made preparations for an experiment to  search for the $\rho_0$  with a simple method (the excitation curve in pion production). Bruno commented: ``Nobody believes in the importance of QED, but I suggest to have more courage and profit of the fact that among the elementary symmetry properties there is C, charge conjugation, which is no less important than P (parity) and T (time inversion). We only know that CPT is in all cases equal to identity, but P is not conserved in weak interactions." Then he mentioned his good friend Rolf Wider\o e, with whom he had discussed, even during war, the possibility of producing 
collisions of oppositely charged particles, moving in opposite directions.
   ``Therefore," he suggested, ``we should immediately modify the synchrotron in order to make two oppositely  rotating beams of electrons against positrons at very large energies to observe such very rare collisions to learn something new." The proposal created some confusion among the listeners, but Salvini intervened immediately to exclude the possibility of modifying the just completed machine. The most intelligent proposal came from Giorgio Ghigo: he said that it would have been much less risky and expensive to build a small prototype to check the effective possibility of matter-antimatter collisions. This is, in simple words, the birth of AdA (Anello di Accumulazione).

A few  Frascati physicists, Corazza, Ghigo and Bernardini, one of the authors of this paper, were immediately enthusiastic. Corazza understood that a very challenging vacuum problem was on the horizon. Ghigo was involved in the magnetic ring design (a kind of cyclotron with betatron orbits at 
{its} periphery), and Bernardini started to work mainly on the problem of injecting two opposite beams in the same ring. He did not feel at ease comparing this scheme to electron-electron colliders in tangent rings, as in the Princeton-Stanford project, appearing much more unnatural. Nevertheless Jerry O'Neill was the Princeton partner of Stanford, fascinated by the momentum transfer in the c. m. system more than by physics.
What must be remembered is that Salvini, Edoardo Amaldi, Felice Ippolito \footnote{ Felice Ippolito was the President of CNEN, the National Committee for Nuclear Energy, at the time the funding agency of the Frascati National Laboratories. } were immediately convinced of the originality of the proposal, and gave all their support to find the money and the approval of INFN, the Italian Institute for Nuclear Physics. 
There were many physicists who  were skeptical about AdA, but Touschek was very firm in giving answers against doubts and in the end he won.
C.B. remembers a peculiar choice of  Bruno when talking with younger physicists: he insisted on the fact that QED was the prototype of every workable quantum field theory; in a sense, this was  the same spirit of the Fermi Lectures at Ann Arbor in 1932, published in the {\it Reviews of Modern Physics}. The same spirit can then be found in Dirac and in the work of Bethe and Jordan. This work has a remote basis that Bruno was continuously mentioning; the harmonic oscillator and itÕs properties. This can be found also in his work on radiative corrections, and in his lectures at the Corsi di Perfezionamento (the graduate school) in Rome. He  also wrote an unpublished note on the vacuum as a dielectric, in which the resonant vacuum was characterized by a peculiar spectrum corresponding to all possible new particles to be produced by supplying the due amount of energy.

Soon after Touschek proposal, in February 1960, a company was selected for the magnet construction, and the order went out.  The magnet was a ring  $160\ cm$ in diameter, with a maximum field of $14500\  Gauss$, with two parts vertically separated by a gap where a  metallic vacuum chamber was inserted. The vacuum system was designed to obtain  a pressure lower than $10^{-7}\  mm \ Hg$, using ionization pumps,  to reduce electron scattering on the residual gas.  There was also space to locate a radio-frequency cavity tuned at $147.2\  MHertz$, corresponding to the second harmonic  of the rotation frequency, with the cavity working at $5.5\  kV$ to compensate for the emission of synchrotron light from circulating particles. 
	
The concept for the injection of particles in the ring was to have a target, internal to  the vacuum chamber, for electron-positron pair production in the ring by an incoming external photon beam. The metallic border of this target would provide, among many others, electron (or positrons) of energy compatible with an oscillating trajectory avoiding the target on successive passages, in order to reach a stable trajectory to continue to circulate in the ring as stored particles. This procedure made the process very inefficient, so that the probability of storing particles in a reasonable amount was very small. Nevertheless, even a tiny group of stored electrons (or positrons) was something which had never been obtained until that time  and a big step forward  \cite{Bernardini:1960zaa}. 
	
\subsection{AdA: from Frascati to Orsay}
\begin{figure}
\centering
\resizebox{0.5\textwidth}{!}{
\frame{\includegraphics{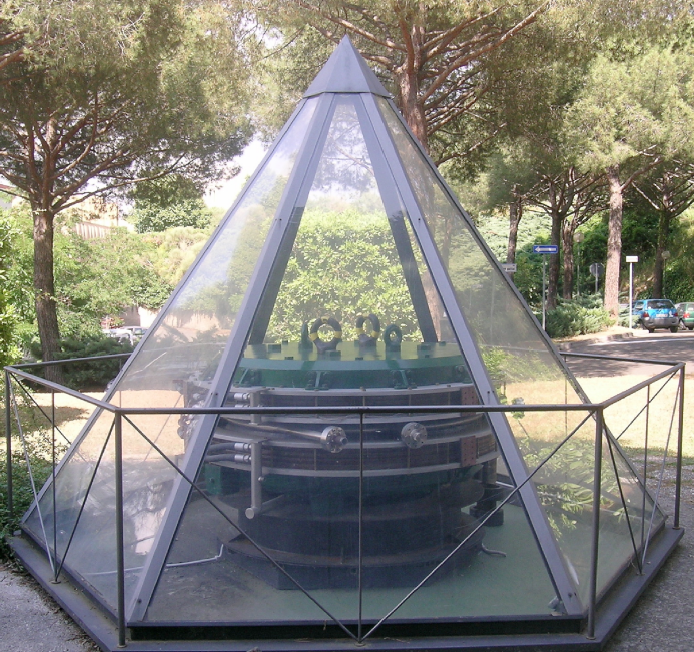}}}
\caption{AdA now, in the grounds of INFN Frascati National Laboratories.}
\label{fig:adanow}
\end{figure}

	In February 1961, just one year after the initial proposal, the AdA group had a clear evidence of the capture of particles on the main orbit, by observing the light emitted by the circulating particles for a time compatible with the beam lifetime estimated from the measured vacuum pressure.  The pressure was still larger than the design value, because of outgassing from the vacuum chamber steel. This was very satisfactory, and AdA always had single electrons stored in the ring to show to astonished  visitors: the visible part of synchrotron light could be seen directly by the human eye \cite{Bernardini:1962zza}. All these preliminary activities were ongoing when  some French colleagues from Orsay, in particular Pierre Marin, visited Frascati. They immediately realized that a well collimated electron beam, such as the one   produced by the Linac at Orsay, would have been a much better injector than the Frascati electron synchrotron. 
	This  visit, in summer 1961, is vividly described by Pierre Marin himself \cite{Marin:2009}, as follows.     \\
{\it Returning from a period at CERN, after my thesis,  I was  searching for my own research direction, and George Bishop suggested that I visit Frascati, where very intriguing things were happening.
 And thus I went there, in the month of August, with Georges Charpak.  In Frascati, a small team of top class physicists showed us with great pride a small machine, in a hall just next to the Frascati electron synchrotron. This was AdA, un vrai bijou, a real jewel.The team, Bruno Touschek, Carlo Bernardini, Giorgio Ghigo, Gianfranco Corazza, Mario Puglisi, Ruggero Querzoli and Giuseppe di Giugno had succeeded in injecting a few thousand
electrons and positrons, accumulating  them  in the AdA ring, circulating for a few hours, until the beams decayed, observing them through the radiation they emitted.\\
 An enormous step for storage rings, but one which strongly needed a larger intensity to observe electron-positron collisions!  }\\
In Fig. \ref{fig:adanow} we show AdA in its present location in Frascati INFN grounds.
	 There had been other ad hoc prepared electron beams attempting to concentrate the pair production on AdA, but these attempts were too inefficient. 
	 AdA was mounted on a Òroasting-jackÓ to invert the field by rotating the magnet, in order to have the ring as near as possible to the synchrotron converter. But this had unavoidable mechanical difficulties with metallic dust  in the vacuum chamber.\footnote{One of us remembers that this effect had been first noticed by Pierre Marin, during his 1961 visit to Frascati. }
	  Therefore the French suggestion   (to use the Linac in Orsay) was considered with extreme interest. 
 	  AdA was transferred to Orsay with a large  truck. Touschek was preoccupied by the difficulties of driving such a heavy vehicle, and he tested the driving problems himself, in the process destroying a street lamp in the Frascati labs. The vacuum pumps were active all the time during the long trip from Frascati to Orsay, to maintain a good vacuum for not less than 48 hours, and reduce the time to start injecting from the Linac.  Ippolito, the president of CNEN, the Italian agency controlling the Frascati Laboratories, asked the Italian Government to intervene when the truck was at the French frontier and the customs officers wanted to control the content of the vacuum chamber! The ring arrived at Orsay in less than 48 hours, and was easily installed in the Salle de cibles 500 MeV. There was immediately a quite remarkable improvement due to the very localized Linac pulse in space and time. Reducing the AdA radio frequency cavity amplitude during the injection pulse produced an unexpected factor 5 in injection efficiency.  The French group included Pierre Marin and Fran\c cois Lacoste and, later, after Lacoste left, Jacques Ha\"issinski, who wrote his {\it Th\`ese d'\'etat} on AdA. 

\subsection{The Touschek effect and start of new $e^+e^-$ projects}
 The next important result was the discovery of a new effect, since then called the ÒTouschek effectÓ.  
The effect is an increase of particle losses, a lifetime reduction, due to large angle electron-electron Coulomb scattering within one bunch.  The scattering transfers transverse into longitudinal momentum. In the laboratory frame the longitudinal momentum change is multiplied by the the relativistic factor, possibly exceeding the acceptance for energy oscillations. 
It is a very sophisticated non-linear effect that Touschek immediately understood. There was  another possible interpretation, much  easier to understand: an excess of outgassing produced by desorption due to an increase of synchrotron radiation on the vacuum chamber wall.  A calculation of the energy dependence of this effect convinced the French and Italian team  that Touschek was right and that, by properly taking it into account, higher energy rings would not suffer as much \cite{Bernardini:1997sc}.  Bernardini recalls that in that night of 1963, when the luminosity would not increase, notwithstanding
the larger electron current, Touschek decided that he needed a quiet place to think and develop the germ of
an idea that had come to his mind. He left to go to the nearby Caf\'e de la Gare d'Orsay and have something to
drink. But, before leaving Touschek said : ``Measure at different energies, and see what happens." When he
came back, a couple of hours later, he said: ``I have understood", and, going to the blackboard, explained in full,
with all the right factors in the formulae, why the intensity would not increase, the cause being 
an
intra-beam effect, which was then called the AdA effect and is now known as the Touschek effect. 

Understanding this and other effects opened the road to higher energy colliders. 
In France with ACO, l'Acc\'el\'erateur Lin\'eaire d'Orsay, and in Italy with  ADONE, new projects were officially started. The Touschek effect was further studied as the {\it multiple Touschek effect} by the Orsay group,  H. Brueck and J. Le Duff \cite{Brueck:1966}. Since that time it has been renamed IBS, Intra Beam Scattering, and is recognized as being important in low emittance storage rings.

News of AdA's results reached immediately the scientific community at large, and accelerated efforts to build electron positron colliders.
Of particular interest was the Dubna Conference of Summer 1963, in the Soviet Union, where  ACO and ADONE projects  were presented.
All the main protagonists of the particle accelerator community participated, among them, from the French-Italian group, Pierre Marin and  one of the writers of this article, C.B..  At the end of the Conference, which was held in Russian, a few selected participants were taken by plane to Novosibirsk. No western scientist had yet seen the new institute. As Pierre Marin remembers : {\it C\'etait une grande pr\'emi\`ere.} And, to everybody's     great surprise,   the scientists saw {\it an electron positron collider VEPPII, of energy 700 MeV, in an advanced stage of construction.} 
VEPPII started operation in 1966.

\subsection{Final measurements}
The last of AdA measurements, which definitely proved  the feasibility of electron-positrons colliders, was a measure of the storage ring luminosity, the rate of events due to beam-beam collisions. It was  found that single electron bremsstrahlung was a good monitor of such events because, even if present in electron-atom collisions, was also produced in electron-positron collisions. Therefore, by measuring the gamma-rays coming from the direction of one of the two counter-rotating beams as a function of the number of the particles of the opposite beam, the results would show a slope on a  constant  background. The slope measures the luminosity, a  procedure which is still used as a luminosity monitor in the present colliders.

The measured height was more than one order of magnitude larger than what one computes by taking into account radiation excitation and damping only Ð assuming there is no vertical dispersion and neglecting the residual gas scattering contribution. This ÔanomalousÕ size could be interpreted as resulting from a small coupling between horizontal and vertical betatron oscillations (it could have arisen from the fringing field at the end of the short straight sections). Later on, such a coupling has played a role in other machines. \\
The final measurement and the conclusions were published  in  \cite{Bernardini:1964lqa}: it was the first time in the world that electron-positron collisions had been obtained in a laboratory  and had been proved to take place.
 \section{Touschek and  ADONE }
ADONE, the electron-positron collider with 3.0 GeV center of mass energy, was envisaged by  Bruno Touschek as soon it became clear that AdA would work. The proposal was sketched in an unofficial note by Touschek, dated November 9th, 1960.  In the note entitled 
{{\fontfamily{cmtt}\selectfont ADONE - A Draft Proposal for a Colliding Beam Experiment},Touschek writes: {\fontfamily{cmtt}\selectfont It is proposed to construct a synchrotron light machine capable of accelerating simultaneously electrons and positrons in identical orbits. The suggested maximum energy is 1.5 GeV for the electrons as the positrons. This energy allows one to produce pairs of all the so called 'elementary particles' so far known, with the exception of the neutrino, which only becomes accessible via the weak interaction channel. }  

The final states mentioned by Touschek are  $2 \gamma$, $ \mu^+ \mu^-$,$ 2 \pi^0$, $ \pi^+\pi^-$, $ K^+ K^-$, $ \ {\bar K}^0 K^0 $, $p {\bar p} $, $n {\bar n}$. Little could Touschek imagine at the time that just above the proposed $3\ GeV$ c.m. energy, Sam Ting and Burton Richter  would discover the particle $J/\Psi$, charmonium, a new state of matter, which opened the way to the Standard Model of elementary particles.

Touschek's note was followed, three months later, by a Frascati Laboratory report, {\it Nota interna n. 68, 27 January 1961}, authored  by  Fernando  Amman, Carlo Bernardini, Raoul  Gatto, Giorgio Ghigo and Bruno Touschek \cite{Amman:1961ad}, entitled ``Anello di Accumulazione per Elettroni. ADONE.".\footnote{``Storage Ring for Electrons. ADONE"}  The text of the proposal is reproduced in \cite{Greco:2004np}. Soon after this note, AdA started to work, and the first electrons were seen to circulate in AdA. Then came the injection problem and the transportation of AdA to Orsay, but ADONE and its construction were the final goal: the Italian team had ADONE in mind, and obstacles were seen as a hurdle to overcome to get to ADONE.

After the completion of the measurements with AdA in Orsay, and the confirmation that collisions had taken places through the observation of the process
\begin{equation}
e^+e^-\rightarrow e^+e^- \gamma,\label{eq:brem}
\end{equation}
the Italian scientists went back to Rome, and the work for ADONE started in earnest. Touschek's interest was on two fronts; on one side he was keen to understand in depth the dynamics of the beams, as we shall see in Sec. ~\ref{sec:touschekacc}, on the other he was concerned about the radiative processes accompanying the collisions and possible difficulties encountered by the planned experiments.  

 The years between 1964 and 1968 show an intense activity at the Frascati Laboratory. ADONE offered the possibility to study the vector bosons, $\rho,\ \omega, \ \phi$, and to hunt for new ones \cite{Bernardini:1965}.  Touschek focused on various problems related to extraction of physics from ADONE. His experience with the betatron had taught him the crucial effect played by radiation emission on the electron's orbit. 
 While Raoul Gatto had put his two students, Guido Altarelli and Franco Buccella, to calculate the rate for process of Eq.~ (\ref{eq:brem}), still  a reference calculation for a number of electron-positron colliders \cite{Altarelli:1964aa}, Touschek assigned the calculation of  double bremsstrahlung to his student Paolo Di Vecchia and one of the young post-docs, Mario Greco. He knew however that at ADONE's energies, no doubt very high energies at the time, the real obstacle would be the summation of many infrared photons.   
\begin{figure}[htb]
\centering
\resizebox{1.0\textwidth}{!}{
\frame{\includegraphics{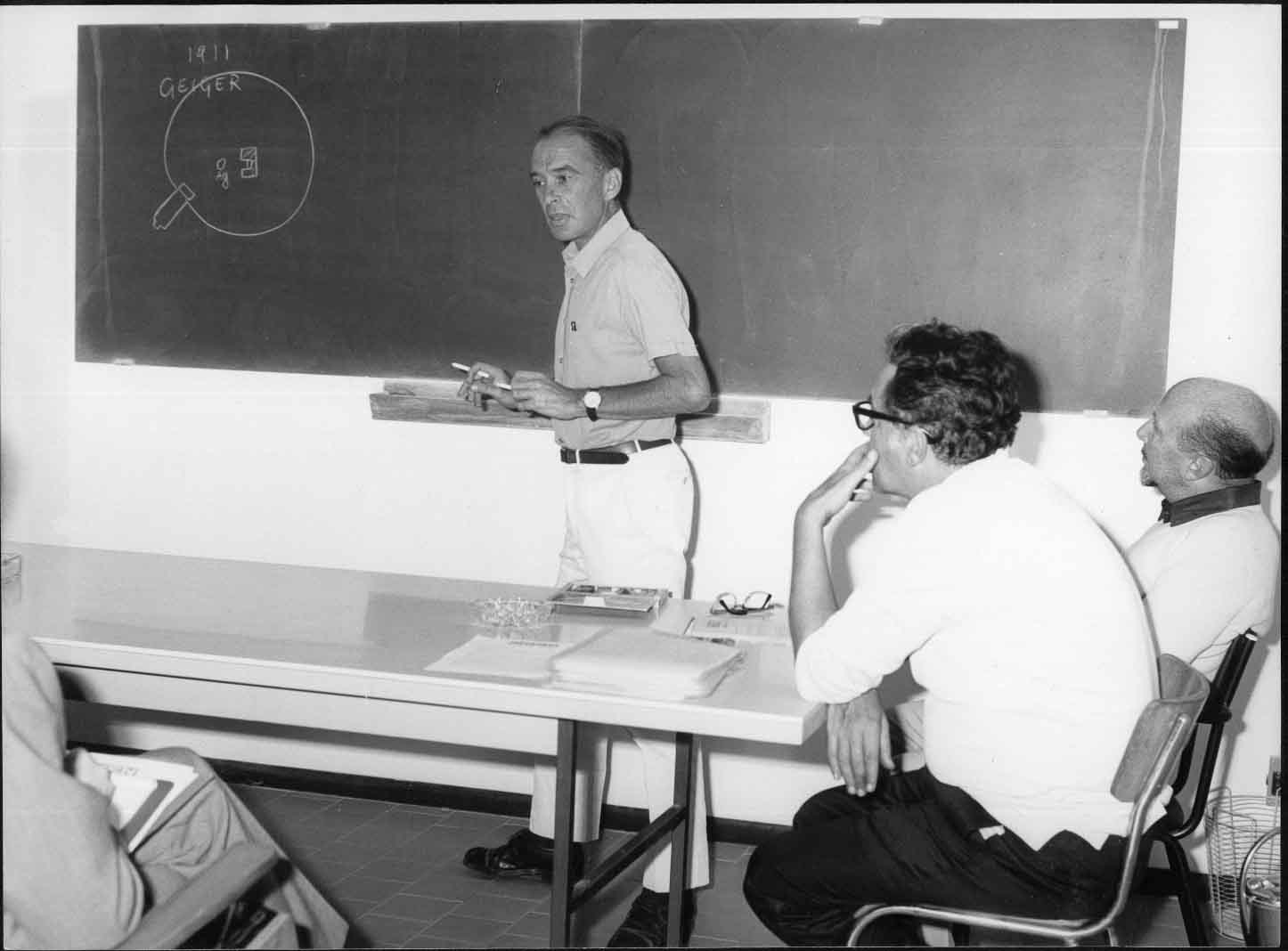}}}
\caption{Touschek in 1964 in the Frascati Laboratory. At right is I.F. Quercia, Laboratories director at the time.}\label{fig:touschekcnen}
\end{figure}
\subsection{1964-74 Radiative Corrections to colliding beam experiments}
When ADONE was well under way, in 1966, Touschek started a program for  what he called the ``administration of radiative corrections".
A young Nigerian student at University of Rome, E. G. Etim, had  asked for a thesis subject, and this work started. After a short internal note, it became clear that the work needed further expansion, and Touschek asked one of the authors of this article, Giulia Pancheri, a newly appointed post-doc in Frascati, in May 1966, to join in the effort. The subject was not very fashionable, it verged mostly on electrodynamics, while new discoveries were at the doors. Many more interesting fields were   coming up at the time, among those of phenomenological interest there were    Current Algebra, Group Theory, Regge Poles. A strong group of young post-docs in theoretical physics, among them Mario Greco, Giancarlo Rossi, Paolo Di Vecchia, Francesco Drago, was at the laboratory at the time. 

One of the authors of this article, G. P., remembers those days as follows: ``Like the other young colleagues, I had indeed started doing some work with current algebra. Thus, at first, Touschek's suggestion to work on radiative corrections to ADONE's experiments was not very attractive.  But one morning, in   June 1966, Touschek wanted to talk to me.  He had a corner office, near the elevator, on the first floor of a building looking down into the small plaza facing the synchrotron. He explained that he wanted us to do the {\it amministrazione delle correzioni radiative}, which he meant  in the English sense, as you administer a medicine, something which you may not like, but you must do, anyway. He spoke  a beautiful Italian, always perfectly  correct in syntax and grammar, but still, and to the very end, with a strong Austrian accent, especially for the sound of the letter {\it r}. He would roll his {\it r}'s in an unforgettable way, which, at the time, we all would  imitate. Still today, the memory of his voice comes back. One  sentence has remained in my mind from that conversation. He said: {\it Signorrina, dobbiamo guadagnarrci il pane e il burrro},\footnote{An extra {\it r} is here added to simulate the way Touschek spoke.} which means {\it Young lady, we must earn our bread and butter}. Actually, the work we did was far from being a boring, straightforward, dry application of already known formulae. He proposed a semi-classical calculation, partly derived from the methods of statistical mechanics, partly perturbative QED calculation, which would, in his word, deal with the 
{\it flood of soft photons which emerge from a high energy collision between elementary particles}." 

The infrared catastrophe problem had been solved in QED by Schwinger and further developed  by Jauch and Rohrlich,  Lomon, Brown and Feynman, Yennie, Frautschi and Suura.

Touschek, in his approach \cite{Etim:1967aa}, stressed  that straightforward perturbation theory does not lend itself easily to deal with this problem, since perturbation theory works in a representation in which the number of photons is diagonal, whereas an experimenter does not see single photons, but rather an unbalance of energy and momentum between the incident and emergent particles. 
The  paper \cite{Etim:1967aa}  is dedicated to obtain realistic expressions for both the energy and the angular distribution of the overall radiation emitted in the process. 
 In this work Touschek   obtained in a very elegant way, the well known expression for the energy loss, i.e.
\begin{equation}
NdP(\omega)=\beta(E)\frac{d\omega}{\omega}(\frac{\omega}{E})^{\beta(E)}
\end{equation}
In conversation, he would jokingly refer to the factor $\beta(E)$ as the {\it Bond factor}, because its value at ADONE could be as large as $ 0.07$. To many of his experimental colleagues, both in Italy, but also in France, at ACO,  and in the Soviet Union, with VEPP2, the simplicity of application and the transparent physical meaning of the expressions he proposed, had a great appeal and the {\it Bond factor} was a joke  remembered for a long time \cite{Greco:2004np}. As we shall describe next, this paper also acquired a particular interest when the $J/\Psi$ was discovered.

One great success of electron-positron accelerators  is  the discovery of the $J/\Psi$ in November 1974 \cite{Augustin:1974xw}, \cite{Aubert:1974js},\cite{Bacci:1974za}. This story has been told many times \cite{Richter:1992eb}, 
 here we shall recall the Frascati contribution to the discovery and the role played by Touschek in  fostering the very precise calculation of radiative corrections, required to extract the $J/\Psi$ width from the data.  On November 11th   1974, a joint press release by SLAC and Brookhaven announced the discovery of a narrow peak in a lepton  pair, as detected in the two different reactions $e^+e^-\rightarrow \mu^+\mu^-/e^+e^-$ at SPEAR, and $pp\rightarrow X + \mu^+\mu^-/e^+e^-$ at a mass  $M_{l^+ l^-}\simeq 3.1 \ GeV$. The news reached Frascati immediately, first from Brookhaven, and then from Stanford. As recalled at the beginning of this section, ADONE   had been built to work only up to a maximum energy of $3\ GeV$, and the $J/\Psi$ was beyond this energy.  However, once the exact energy  of the resonance peak was communicated from SPEAR, ADONE in two days reached it  and, to everybody's great excitement, the counters started accumulating data at a rate never seen before. 

In the days and months to follow, two were the main goals of all physicists working at SPEAR and ADONE, namely to define with precision the width of the particle and to discover the higher energies recurrences, which would establish beyond doubt that it was a quarkonium state. In Frascati, the second goal was out of the question, the first was attainable. But, for this, one needed Touschek' s precise radiative corrections methods. Indeed, once more, it was during a meeting held in Frascati to discuss the collection of data, that Touschek said to the young collaborators he had trained in this techniques a few years before: ``You must do the radiative corrections".   It is thanks to Touschek, that a serious effort was undertaken by Greco, Pancheri and Yogendra Srivastava, to develop    the precise calculation of resummed soft photon emission in presence of a very narrow resonance and  bring it  to a successful conclusion \cite{Greco:1975rm}. Later, when Touschek saw the theoretical curve applied to the Frascati data, being absolutely confident on the method used to obtain the curve,   he observed: ``The errors to the data are too large!". We all felt this was a great compliment to our work, and   indeed the errors were too large and were later reduced.  We show in Fig.~\ref{fig:catching data} two of Touschek's drawings possibly related to this period and the excitement about the $J/\Psi$ discovery.
\begin{figure}
\resizebox{1.0\textwidth}{!}{
\frame{\includegraphics{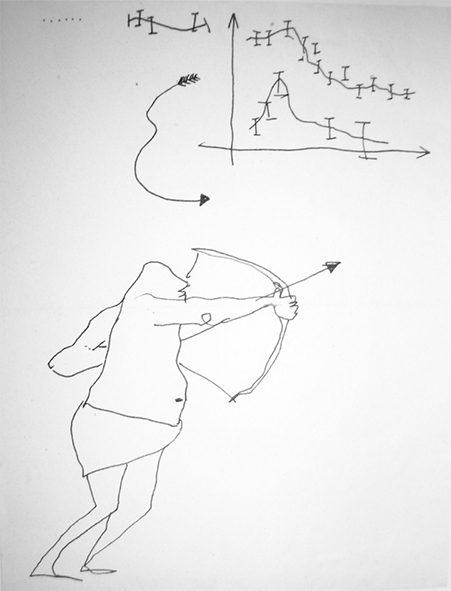}}
\hspace{0.5cm}
\frame{\includegraphics{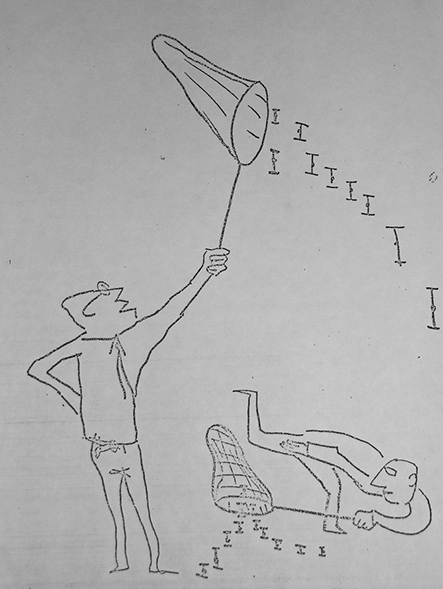}}}
\caption{Two drawings by Touschek, around the time of the discovery of the $J/\Psi$. The radiative tail of the resonance is clearly seen in both drawings.}
\label{fig:catching data}
\end{figure}

 There is no doubt that Touschek's contribution to the field of electron-positron collisions goes beyond proposing AdA and ADONE, and that his influence on the theorists at Frascati and University of Rome has been larger than what may appear from the literature.

\subsection{Contributions to ADONE accelerator physics}\label{sec:touschekacc}
Touschek's contribution to accelerator science, and in particular to ADONE, goes beyond the initial proposal and what is recorded in the literature and we will try in this section to discuss both the published and unpublished work.

ADONE was designed during the 1960s to reach what, at that time, was a very high luminosity, larger than $10^{30} /cm^2 s$. The high luminosity was critically important to carrying out its mission: the exploration of high-energy physics, including strong interactions, in a new way. It was a project pushing the limits on accelerator physics and technology.  There were many uncertainties in the design, even if AdA had already provided some new discoveries and information. The design luminosity required a very large electron and positron stored current, 100 mA per beam. This immediately raised the issue of the stability of such large accumulated currents, the problem of the beam Ðbeam interaction at collision points and the limitations it imposed on the particle density in the two beams, the beam lifetime, including the residual gas effects, ion trapping effects, which had been observed in AdA, and the Touscheck effect. The Touschek effect \cite{Bernardini:1997sc} in ADONE was calculated using his theory extended to include relativistic effects. Two high intensity instabilities, the negative mass instability \cite{Nielsen:1994sr}
 and the resistive wall instability \cite{Laslett:1965ri}
 had been observed and analyzed in other accelerators. The theoretical analysis had been established for the case of coasting beams of interest to proton accelerators and colliders. These theories had to be extended to the case of bunched electron-positron beams like those in ADONE. The extension of the resistive wall instability to bunched beams was done by Courant and Sessler \cite{Courant:1966rx} and, at Frascati, by Touschek with two young coworkers, Ferlenghi and Pellegrini \cite{Ferlenghi:1966}. The last is one of the authors of this paper and he remembers quite well the excitement of working with Bruno. What he learned from him during that period had a deep impact on all his subsequent work on collective instabilities in high intensity electron beams.
The expectation that some new beam physics would be found when injecting beams in ADONE was well justified. It was soon clear that collective instabilities were playing a major role, in fact limiting initially the stored current to about $100\  \mu A$, against the 100 mA of the design.
The situation was well summarized by Amman in a paper presented at the 1969 Particle Accelerator conference \cite{Amman:1969gt}:
 {\it It may seem strange that eight years after the initial operation of a storage ring, only one $e^-e^-$ (the Princeton-Stanford 550 MeV) and two $e^+e^-$ rings, VEPP-II and ACO, have produced high energy physics results, and these are limited to experiments with very high cross section I would like to remark that the first beam instabilities observed on the Princeton-Stanford $e^-e^-$ ring, and interpreted as being due to the resistance of the walls, opened a new era in the accelerator field: it has been realized for the first time that the interaction of the beam with its environment makes a circular accelerator an essentially unstable system, that can become stable, in virtue of the Landau damping, when the beam density is not too high and the non linearities in the focusing forces give a frequency distribution of the particles large enough to compete with the instabilities.}

During these days, when trying to understand what was the problem, Touschek's  presence in the control room and the discussion with him of the effects being seen was crucially important. It was soon established that the main problems were transverse and longitudinal instabilities. The longitudinal effects were attributed to resonances in the RF cavity, driving large longitudinal oscillations at low beam current. Together with Littauer, Sands and one of us, Touschek was  very active in  analyzing this problem, which  was solved in the end by redesigning the RF cavity, and damping the most damaging resonances.

The transverse instabilities did not fit the bunched beam resistive wall theory. The discussion with Touschek and Matthew Sands, who was also spending time at Frascati to work on ADONE, led to identify the presence of high frequency effects, corresponding to modes internal to each bunch, and driven mostly by the clearing electrodes installed to eliminate ion trapping \cite{pellegrini69}. Once understood the source of the new effect solutions were found to control them and increase the current to the design value.

 We have outlined the attention paid by Touschek to collective instabilities, but this is only one of the fundamental problems in  accelerator physics which Touschek studied.  One of the problems arising in the stability of  ultra relativistic beams of electrons and positrons is, of course,  the Touschek's effect, discovered in AdA, in 1963 \cite{Bernardini:1997sc}. But, even before AdA, together with Carlo Bernardini, Touschek had addressed the problem of beam losses
  in 1960 \cite{Bernardini:1960ad}. Later, when ADONE was in an advanced stage of  construction,  he again studied  this problem, addressing the beam instability in a storage ring \cite{Ferlenghi:1966} due to the chamber  wall's resistivity. Touschek's contribution to solving a number of problems which could reduce the accelerator's luminosity are also recalled by Fernando Amman in a letter to Amaldi, after Touschek's death. Amman, who directed the construction of ADONE and co-authored the formal proposal to build ADONE in 1961 \cite{Amman:1961ad},   writes \cite{Amaldi:1981be}: {\it Bruno lived intensely the various stages of ADONE: the design as well as the construction period (1965-67) and especially the difficult year 1968 during which various beam instabilities were studied and  cured. \dots he felt that this was really the materialization of his original idea. Whenever a new problem came up on which he was certain to be able to contribute, Bruno was present, without the need to look for him; an example was the case of the transverse instabilities (1965) \dots}

\section{Touschek's sense of humor:  drawings and anecdotes}
Touschek had a keen sense of humor and a distinctive ability to render it graphically. He had inherited his capacity to draw from his mother's family, active, as mentioned, in the Viennese artistic circles of early 1900's. A glimpse of these ties can be seen from of  one of his letters home, during the hard war years, when food was getting scarce, and his loneliness deepened: in the letter, he asked the father to send him  two paintings which had adorned his room back home and which he wanted in Berlin  to relieve his homesickness: one by his uncle Oskar Weltmann, the other  by Edgar Schiele, the great Viennese painter.

 Touschek's drawings have survived both the destructions of war and the carelessness of their author. The letters home, graphically illustrating some   daily occurrences,  were carefully preserved by his father, and are now with Touschek's son. The drawings from these letters are in  a small number, however, and lack some of the caustic humor of Touschek's later years, particularly those referring to  the 1968 student unrest in Rome or university life, which span a period from 1952 to, perhaps, 1976, and constitute a unique commentary  to the  years before, during and after the 1968 student unrest, at University of Rome. The drawings are   part of a large body  of drawings  which he would just leave casually around. His friends   would pick them up, and cherish them. After his death, many were published  in Amaldi's biography \cite{Amaldi:1981be} and some can also be found in \cite{Bonolis:2011wa}. A good number  of them} were also preserved by  his wife, Mrs. Elspeth Touschek:  finding  the crumpled pieces of paper  in   Touschek's pockets, while looking after his clothes, she would iron out the papers   and keep them in some boxes.\footnote{Private communication by E. Touschek to Luisa Bonolis  and G.P.}

  In addition to the drawings,  anecdotes about him abound in the memory of his friends. 
 During the period of construction of the synchrotron, there is the story of the motorcycle accident, whereupon Touschek was sent to the hospital in Frascati, with a slight concussion, after hitting the back of a truck.  Since Touschek appeared confused, he was sent to the neurological and psychiatric ward.  The director of the  ward went to look at him and inquire whom he was, what had happened. Touschek, perhaps still under shock, and in his peculiar italian, with a strong Austrian accent, said that if you perform  a {\it time inversion}, it could be said that he had been hit by a truck doing reverse maneuvering. He then  added  that he was the director of the graduate program in physics at University of Rome. The whole story appeared improbable to the doctor visiting him, and he ordered the patient to  receive electroshock therapy.   At this point, a young intern, later to become a well known neurobiologist, Valentino Braitenberg, stopped the procedure saying: {\it Don't! He could be a theoretical physicistÉ}.\footnote{Recollections by   C.B., in the docu-film {\it Bruno Touschek and the art of Physics, INFN 2003}.}
 
 Through his family background, and his many adventurous life changes, Touschek acquired a vast inter-european culture, which often surprised his students and colleagues. He would make his young Italian colleagues listen to recordings of songs from  {\it H.M.S. Pinafore}: at the time not only it was difficult to find young people versed enough in the language to understand the songs, but the humor was also very far from the spirit of the time, even in a capital city such as Rome.
 
Bruno's capacity to see satirical connections, is  reflected in  an anectode, related by Carlo Bernardini concerning  an evening at the Touscheks', during a dark period for the Frascati Laboratory, the so-called Ippolito trial. 
 It was an evening among some close friends of Felice Ippolito, the President of CNEN,  the funding agency for the Laboratories at the time. Ippolito had been accused of misuse of funds. The get together had been  meant to discuss  a strategy to clear Ippolito of such accusation and show that his  behavior amounted to no more than  an attempt to cut through  the many  obstacles placed   by Italian bureaucracy to an efficient running of research programs.   But then, as the evening went by, the conversation  got stranded  and turned to more trivial gossip and chatter. At one point, Touschek lost patience since the issue was of burning importance, and he could not stand the levity of the conversation. Turning to Salvini, he asked: ``By the way Mrs. Lincoln, how was the play? ".\footnote{In Italian, he said: ``A parte cio',  Signora Lincoln, com'era lo spettacolo,?'' } This sentence became a frequent joke among Touschek's friends.
 
 Most of the anecdotes refer to his use of the Italian language in a very humorous way, and they are lost in translation, but one in English, reflecting how he liked to distort words, refers to a trip he took  to England. He was carrying some cigars, in fact 90 of them. He put 71 at the bottom of his suitcase, under books and clothes, and 19 in plain view. At the frontier, the custom officer asked: \\
 "What are you carrying? "\\
 Touschek answered: "Ninety cigars".\\
 "You mean  nineteen?", said the custom officer.\\
 "No" insisted Bruno," ninety".\\
 "It is nineteen, go,  just goÉ" said the custom officer, clearly annoyed and also piteous of Touschek's poor English, and so Bruno passed customs with   ninety cigars, in his suitcase.
 
 Another time, in Paris for a seminar (perhaps when he had been invited in Spring 1962 to finalize AdA's transport from Frascati to Orsay), he lost his shoes. The anecdote is told by the theoretical physicist Maurice L\'evy, who was his host in Paris.\footnote{Private communication to G.P. in May 2013.} When L\'evy arrived at the hotel to take Bruno to the University, he found him in great agitation. Bruno's shoes, which he had put outside his room the night before to be cleaned (as  was the custom in Italy), had disappeared and he had no other pair. Inquiries with the hotel staff, with Bruno more and more upset, led nowhere. Finally, L\'evy and Bruno, barefoot, walked to a shoe shop somewhere in the Champs Elis\'ees, a  new pair was purchased and the visit could continue. \\

\begin{figure}
\resizebox{0.325\textwidth}{!}{
\frame{\includegraphics{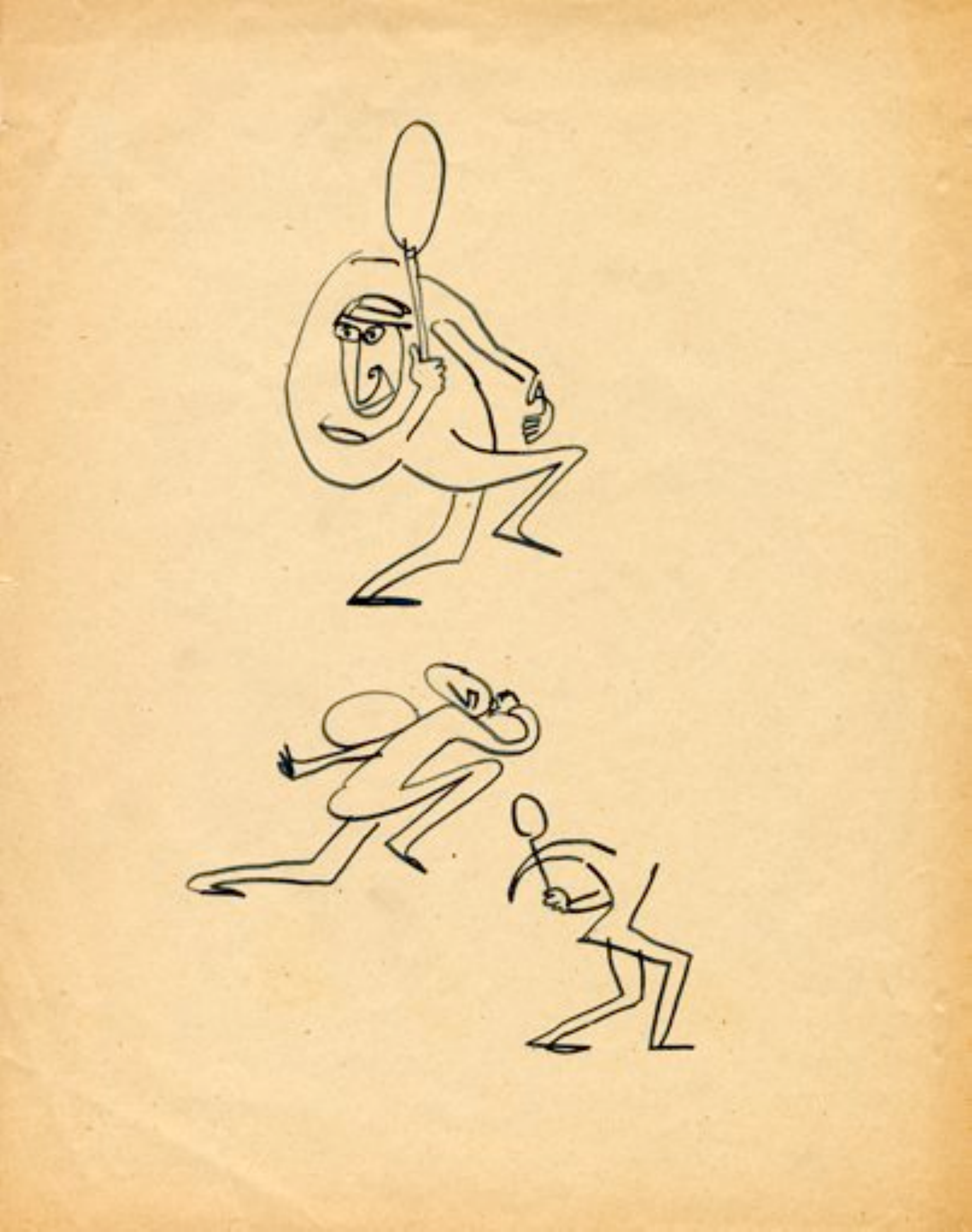}}}
\hspace{1cm}
\resizebox{0.5\textwidth}{!}{
\frame{\includegraphics{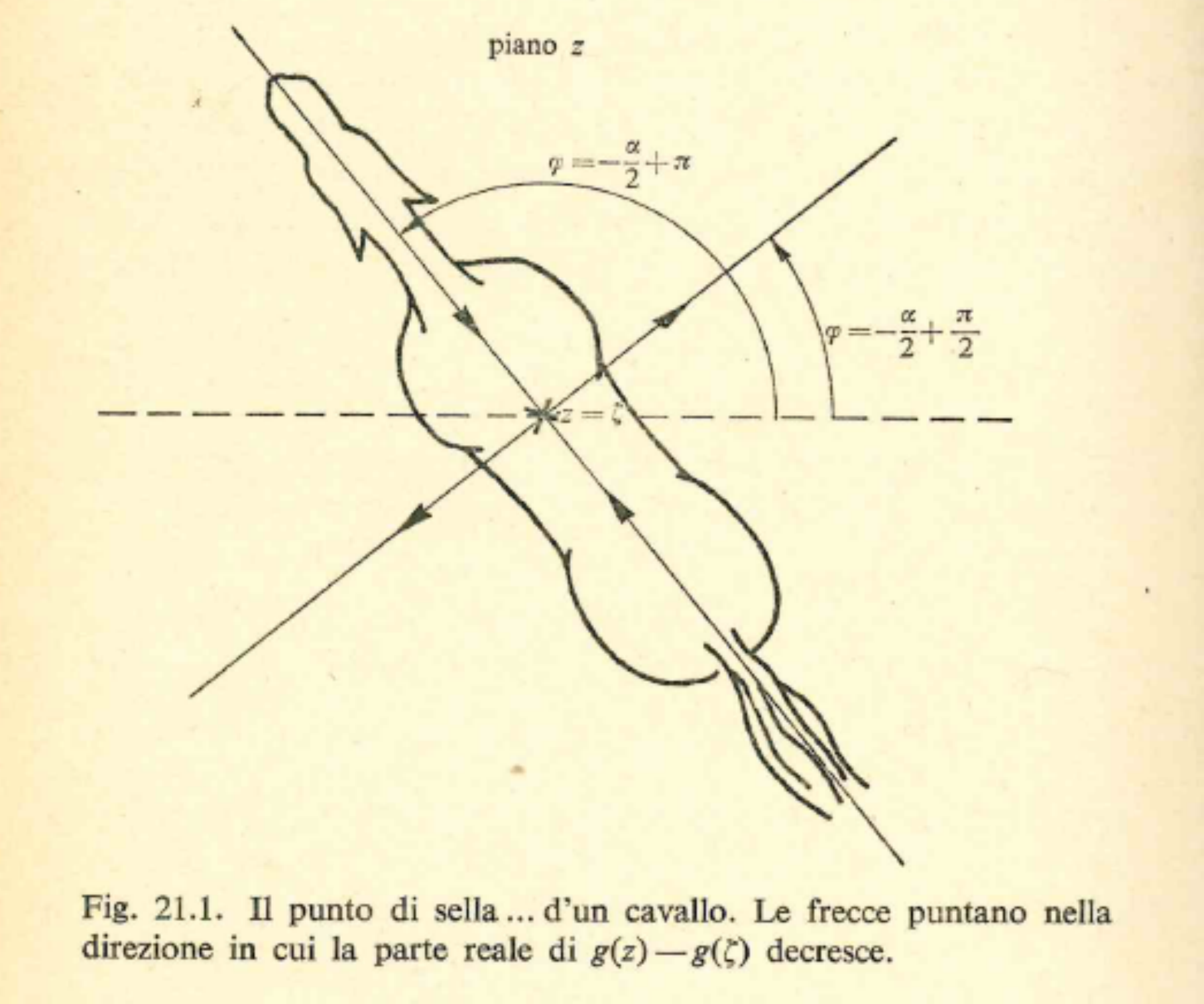}}}
\caption{Two drawings by Bruno Touschek:  tennis players at left, and, at right,  an interpretation of the saddle point method, in the book {\it Meccanica statistica} by B. Touschek and G.C. Rossi, 1970 Ed. Boringhieri, Italy.}\label{fig:tennis}
\end{figure}

Before leaving our  description of Touschek and his life, we shall show two more of Touschek's drawings in Fig. ~\ref{fig:tennis}. The one at left is  from his early period in Rome, and recalls how he  used to enjoy playing tennis with his Roman colleagues, Edoardo Amaldi and Francesco Calogero. The drawing at right, is from his more mature period and illustrates the saddle-point method. He is remembered as drawing a saddle one day, and then add: ``This is the saddle, and  now I shall draw the horse as well."

\section{Final remarks}
Touschek's life developed across many European countries and through the Second World War.  It  is very clearly and exhaustively presented in Amaldi's biography, and completed from a reading of Touschek's letters to his parents. From   E. Amaldi's work,
one learns,  in particular,   about  Touschek's apparently strange university career in Italy. This is due to the fact that Touschek remained an Austrian citizen to the end of his life, and that Italian law did not allow foreigners to hold regular professorship positions in the state universities, such as the University of Rome. After a change of the law, in 
1973, Touschek was appointed Extraordinary Professor, a position which would  become that of Ordinary (Full) Professor after three years, upon approval by the Ministry of Education, a standard practice. But the approval  required submission of a number of documents and certificates. Touschek refused to submit the application, considering it  unbearably burdensome and below his dignity, as if he had to prove, once more, that he deserved to be confirmed in the post. Finally, his friends did it for him. Alas, when  he was finally appointed  full (Ordinary) professor in early 1978,  it was too late for him to enjoy the position he should have held so many years before. On May 25th, 1978, Bruno died in Innsbruck, in Austria, the country he was still a citizen of, and that he had left as a twenty one year old,  to begin his great adventure in science.\\

\section*{Acknowledgments}
We are thankful to Luisa Bonolis, who has given us permission to use yet unpublished material she has been discovering and recording. We are also grateful for her  translation from German of parts  of  Touschek's letters to his family, and for a critical reading of this manuscript. We thank  Francis Touschek for allowing publication of excerpts from his father's letters and drawings. Collaboration and help from  the Information Service of INFN Frascati National Laboratories are gratefully acknowledged. We thank Andrea Ghigo for retrieving unpublished ADONE notes, and O. Ciaffoni for technical help with the figures.  G.P. acknowledges hospitality at MIT/CTP during completion of this manuscript. Finally, we are indebted to Jacques Ha\"issinski for advice, and for supplying   bibliographic material and copies of  letters exchanged between LAL, Frascati and University of Rome.
\bibliography{touschek-rast}
\end{document}